\documentclass[journal]{IEEEtran}
\usepackage{cite}

\ifCLASSINFOpdf
\else
\fi
\usepackage{bm}
\usepackage{amsmath,amssymb}

\usepackage{mathrsfs} 
\usepackage{xcolor}

\usepackage{float}   
\usepackage[ short ]{ optidef }

\usepackage{kotex}
\usepackage{comment}
\usepackage{bbm}

\usepackage{algorithm}
\usepackage{algorithmicx}
\usepackage{algpseudocode}

\usepackage{cuted}

\usepackage{graphicx}

\usepackage{makecell}
\usepackage[table]{xcolor}
\definecolor{LightBlue}{RGB}{235,242,255}
\definecolor{LightGray}{RGB}{242,242,242}
\definecolor{SectionGray}{RGB}{225,225,225}
\definecolor{sb}{RGB}{0, 30, 255}

\usepackage{amsthm}
\theoremstyle{remark}
\newtheorem{remark}{Remark}

\usepackage{enumitem}

\begin{document}
%
\title{Demand-Aware Cooperative Transmission Design 
for Energy-Efficient LEO Satellite Networks}
%
%
%

\author{Wooseok Cha,~\IEEEmembership{Graduate Student Member,~IEEE},~Kyeongsoo Kim,~\IEEEmembership{Graduate Student Member,~IEEE},\\Seonghoon Kim,~\IEEEmembership{Graduate Student Member,~IEEE},~Junil~Choi,~\IEEEmembership{Senior Member,~IEEE},\\and Jihwan~P.~Choi,~\IEEEmembership{Senior Member,~IEEE}
        \thanks{The work of Wooseok Cha and and Junil Choi was supported by Korea Research Institute for defense Technology planning and advancement(KRIT) grant funded by the Korea government(DAPA(Defense Acquisition Program Administration)) (KRIT-CT-22-040, Heterogeneous Satellite constellation based ISR Research Center, 2022), and in part by the Institute of Information $\&$ Communications Technology Planning $\&$ Evaluation(IITP)-ITRC(Information Technology Research Center) grant funded by the Korea government(MSIT)(IITP-2026-RS-2020-II201787, contribution rate $20\%$). The work of Kyeongsoo Kim, Seonghoon Kim, and Jihwan P. Choi was supported by the Korea Institute for Advancement of Technology (KIAT) grant funded by the Korea Government (MOTIE): P0028333 and the Materials Components Technology Development Program (RS-2024-00452151) funded by the Ministry of Trade, Industry $\&$ Energy (MOTIE, Korea). Wooseok Cha and Kyeongsoo Kim contributed equally to this work. (Corresponding authors: Junil Choi; Jihwan P. Choi.)}
        \vspace{-0.5cm}
        \thanks{W. Cha, and J. Choi are with the School of Electrical Engineering, KAIST, Daejeon 34141, Republic of Korea. (email: wscha, junil@kaist.ac.kr)}%
        \thanks{K. Kim, S. Kim, and J. P. Choi are with the Department of Aerospace Engineering, KAIST, Daejeon 34141, Republic of Korea. (email: shtk125, seonghoon, jhch
        @kaist.ac.kr)}%
}

%
%

\markboth{}%
{Shell \MakeLowercase{\textit{et al.}}: Bare Demo of IEEEtran.cls for IEEE Journals}
%



\maketitle

\begin{abstract}
Low Earth orbit (LEO) satellite networks are envisioned as a promising solution for providing ubiquitous connectivity and narrowing the digital divide. The extensive footprint of LEO satellite constellations enables broad coverage, resulting in spatially non-uniform traffic demand across the serviced areas. Meanwhile, stringent on-board power constraints make power-intensive transmission architectures less attractive and motivate energy-efficient transmission strategies that effectively exploit scarce satellite network resources. To this end, this paper proposes a cooperative transmission framework that jointly accounts for non-uniform traffic demand and network-wide power consumption. Each LEO satellite integrates hybrid precoding (HPC), radio frequency (RF) chain activation, and hardware quantization, while user-equipment (UE)-centric satellite clusters are organized using statistical channel state information (sCSI) and traffic demands. A framework for joint optimization of cooperative transmission architecture and resource allocation is designed to maximize demand-aware energy efficiency (EE), resulting in a mixed-integer nonlinear program (MINLP) for which finding a globally optimal solution is generally intractable. Accordingly, a two-stage algorithm is developed under a distributed linear precoding structure, in which a modified cross-entropy (CE) method searches over discrete variables, while fractional programming is employed for transmit power allocation. Numerical results indicate that the proposed framework outperforms benchmark schemes while accounting for traffic demands and EE.
\end{abstract}


%
\IEEEpeerreviewmaketitle

\section{Introduction}
%
%
%
%
\IEEEPARstart{N}{on}-terrestrial networks (NTNs) are emerging as a next-generation infrastructure to reduce the digital divide and provide ubiquitous connectivity through wide coverage and flexible deployment\cite{Intro1}. In particular, low Earth orbit (LEO) satellites have attracted growing attention because their low altitudes reduce latency and enable higher achievable data rates\cite{LiYou_1,Li_hybrid,Liu2024,ratematching,kimcellfree,ZhangISL}. Moreover, advances in launch technologies, together with improved satellite manufacturing and integration\cite{jungdonghyun}, have reduced deployment costs, thereby facilitating massive deployment of LEO satellite constellations. These technological advances create favorable conditions for deploying and operating a large number of satellites, and progressively broaden the practicality of leveraging multi-LEO satellite networks across diverse scenarios. 
Meanwhile, in terrestrial wireless communications, multiple-input multiple-output (MIMO) has long been a cornerstone technique to enhance spectral efficiency (SE), and substantial studies have been devoted to MIMO and associated transceiver designs\cite{Choi2022,9238423,Li2015,9064545,6457363}. These developments have also been extended to satellite communications, where MIMO for LEO satellite systems is being actively studied to improve link performance\cite{LiYou_1,Li_hybrid,Liu2024,ratematching,kimcellfree,ZhangISL}. In multi-LEO satellite networks, the concurrent visibility of multiple satellites provides additional spatial diversity and transmission flexibility. Consequently, multi-antenna, multi-satellite transmission is expected to be a key enabler for achieving high performance and efficient resource utilization.\\
\indent Despite these advances, LEO satellite networks still pose practical design challenges. Wide-area coverage inherently leads to an interference management problem\cite{jeffrey,seunggi} as well as spatially non-uniform traffic demands\cite{1545873,ratematching,ZhaoDemand2025,seunggi}, and the network is expected to continuously accommodate such demands under limited resources. Furthermore, satellite payloads operate under tight power and hardware budgets, which can limit the use of advanced transceiver architectures that entail high power consumption and hardware complexity. Consequently, transmission strategies for LEO satellite networks necessitate going beyond the conventional objective of SE maximization alone, by accommodating non-uniform traffic demands while accounting for power consumption and implementation complexity, especially in the context of massive MIMO-based architectures\cite{LiYou_1}.

\vspace{-0.1cm}\subsection{Related Work}

The stringent power constraints of LEO satellites have motivated extensive research on energy efficiency (EE) in transmission design\cite{Liu2024,9238423}, including studies that incorporate circuit power models accounting for the quantization levels of digital-to-analog converters (DACs) and phase shifters (PSs) \cite{Choi2022,Li_hybrid}. 
In addition, hybrid precoding (HPC) has been explored as an energy-efficient and practical transceiver architecture \cite{Li_hybrid,Liu2024}, since conventional fully digital precoding requires a separate radio frequency (RF) chain for each antenna, often making it impractical for LEO satellite systems due to its high power consumption and complexity.
Moreover, various methods have been proposed to improve EE, including RF chain or antenna activation \cite{9238423,Choi2022} and neural network-based precoding optimization \cite{ZhouGNN2025}.

Cooperative transmission has been investigated as an effective approach to enhance communication performance in wireless networks \cite{WMMF,9064545}. In the context of LEO satellite networks, such cooperation has been studied by leveraging multi-satellite visibility within a time window to enable joint transmission from multiple satellites, moving beyond satellite-centric operation, in which each user is served by a single satellite \cite{kimcellfree,ZhangISL,jungdonghyun,sujuwuin}. The cooperative operation inherently provides macro-diversity and improved link reliability, while existing studies have investigated dynamic coordinated beamforming for coverage enhancement and cooperative transmission under hardware quantization distortions \cite{KimCoordinated2024,Kim2022}.

To handle non-uniform traffic in wireless networks, numerous performance metrics have been studied. In particular, weighted max–min fairness (wMMF) balances demand satisfaction across user-equipments (UEs) to enhance fairness \cite{WMMF}, while rate-matching approaches minimize the gap between each traffic demand and the achievable rate \cite{ratematching}. These metrics have been widely used to evaluate traffic demand satisfaction under different design objectives in both terrestrial and satellite communication systems.

\vspace{-0.2cm}\subsection{Contributions}
The distinctive aspect of this work lies in jointly integrating energy-efficient precoding, cooperative transmission, and demand-aware optimization into a unified framework for multi-LEO satellite networks, whereas existing studies have mostly considered these aspects separately.
In particular, this paper focuses on the interplay between non-uniform traffic demands and overall power consumption, and develops a holistic transmission design framework that integrates cooperative transmission with comprehensive power consumption modeling. Specifically, the downlink (DL) cooperative transmission in LEO satellite networks is investigated, where HPC vectors, RF chain activation, UE association, and transmit power allocation are jointly optimized to maximize demand-aware EE, which simultaneously accounts for traffic demand satisfaction and overall energy consumption. The main contributions of this paper are summarized as follows.

\begin{itemize}
    \item A cooperative transmission design is considered for an LEO satellite network that supports both fully-connected (FC) and partially-connected (PC) HPC architectures. DAC quantization is modeled using the additive quantization noise model (AQNM), and PSs are modeled with finite phase resolution, thereby capturing realistic hardware characteristics and energy consumption. Both circuit and transmit power consumptions are included, ensuring that variations in the design variables directly affect the total network power consumption. Consistent with recent work in satellite communications, statistical channel state information (sCSI), which varies slowly as it captures long-term channel characteristics, is employed, from which a lower bound on the ergodic rate is obtained. Subsequently, a demand-aware EE metric is defined as the ratio between the minimum demand-normalized rate, where each UE rate is divided by its respective traffic demand, and the overall network power consumption.
    \item A transmission framework with centralized optimization and distributed precoding is developed to obtain a high-quality solution for the resulting mixed-integer nonlinear programming (MINLP) in a tractable manner and to reduce the signaling overhead between the gateway and satellites in large-array multi-satellite cooperative systems. At the gateway, the central processing unit (CPU) performs optimization in two stages over low-dimensional variables: RF chain activation and UE association are optimized using a modified cross-entropy (CE) method that iteratively samples configurations and scores each sample under a simplified common power level optimization, and transmit power is then optimized by integrating the Dinkelbach method with a quadratic transform. In addition, leveraging the single angle-of-departure (AoD) nature of satellite channels for each UE, an sCSI-based linear precoding design is adopted, where each satellite computes its precoder locally without large-scale information exchange.
    \item Numerical results demonstrate that the proposed framework consistently achieves higher demand-aware EE than benchmark schemes across a broad range of system parameters. The impacts of hardware quantization resolutions and imperfect pre-compensation for Doppler shifts and propagation delays are also analyzed.
    Moreover, comparisons with conventional sum-rate-based EE metrics show that the demand-aware EE metric satisfies UE traffic demands more fairly while effectively utilizing available system resources. Convergence is analyzed for each stage of the proposed two-stage iterative algorithm performed at the gateway, and both stages are shown to converge reliably.
\end{itemize}   

The remainder of this paper is organized as follows. Section~\ref{section2} details the system model for the proposed LEO satellite cooperative transmission design and defines the performance metrics. Section~\ref{section3} formulates the optimization problem. Section~\ref{section4} develops the proposed algorithms. Section~\ref{section5} reports numerical results. Section~\ref{section6} concludes the work.

\textit{Notation:} Vectors and matrices are denoted by bold lowercase and uppercase letters, respectively, while sets are denoted by calligraphic letters. The sets of non-negative integers, real numbers, and complex numbers are denoted by $\mathbb{Z}_{+}$, $\mathbb{R}$, and $\mathbb{C}$, respectively, and $\mathbb{C}^m$ denotes the $m$-th Cartesian power of $\mathbb{C}$. The $m \times m$ identity matrix is denoted by $\mathbf{I}_m$. The $(i,j)$-th element of a matrix $\mathbf{X}$ is denoted by $[\mathbf{X}]_{i,j}$, and the $i$-th element of a vector $\mathbf{x}$ by $[\mathbf{x}]_i$. The transpose and Hermitian transpose are denoted by $(\cdot)^\top$ and $(\cdot)^{\mathrm{H}}$, respectively. The phase of a complex scalar is denoted by $\angle(\cdot)$. A zero-mean, circularly symmetric complex Gaussian random variable with variance $\sigma^2$ is denoted by $\mathcal{CN}(0,\sigma^2)$. Expectation and variance are denoted by $\mathbb{E}\{\cdot\}$ and $\mathbb{V}\{\cdot\}$, respectively. The trace operator is denoted by $\mathrm{tr}(\cdot)$ and the cardinality of a set by $|\cdot|$. The operator $\mathrm{diag}(\cdot)$ forms a diagonal matrix from its vector argument; for a square matrix, it keeps only the diagonal entries and sets all the others to zero. The Kronecker product is denoted by $\otimes$. $\|\cdot\|_0$ denotes the $\ell_0$-norm, while $\|\cdot\|$ denotes the $\ell_2$-norm. $\mathcal{U}_{[a,b]}$ denotes the uniform distribution over $[a,b]$.

\begin{table}[t]
\centering
\caption{Summary of Key Notation}
\label{tab:key_notation} 
\renewcommand{\arraystretch}{1.3}
\begin{tabular}{c|p{0.62\columnwidth}}
\Xhline{0.8pt}
Notation & Definition \\
\hline
\rowcolor{SectionGray}
\multicolumn{2}{l}{System model and channel notation} \\
\hline
\rowcolor{LightGray}
$L$, $K$ & Numbers of LEO satellites and UEs\\
$\mathcal{L}$,\,$\mathcal{L}_k$ & Sets of all satellites and satellites serving UE $k$\\
\rowcolor{LightGray}
$\mathcal{K}$,\,$\mathcal{K}_\ell$ & Sets of all UEs and UEs served by satellite $\ell$ \\
$\varsigma\!\in\!
\{\mathrm{FC},\!\mathrm{PC}\}$ & HPC architecture index \\
\rowcolor{LightGray}
$N_\mathrm{t},\,N_\mathrm{RF},N^\varsigma_{\mathrm{PS}}$ & Numbers of antennas, RF chains, and PSs\\
$\mathbf{h}_{\ell,k}$ & Channel vector from satellite $\ell$ to UE $k$ \\
\rowcolor{LightGray}
$\nu^\mathrm{(sat)}_{\ell,k},\tau^\mathrm{(min)}_{\ell,k}$ & Satellite Doppler shift and minimum propagation delay \\ 
$\mathbf a(\theta,\phi)\,\text{(or } \mathbf{a}_{\ell,k} \text{)}$ & Array response vector with AoDs \\
\rowcolor{LightGray}
$\gamma_{\ell,k},\kappa_{\ell,k},\upsilon_{\ell,k}$ & Path loss, Rician K-factor, and scaled path loss\\
$\alpha_{\ell,k}\!\sim\!\mathcal{CN}(0,1)$ & Small-scale fading coefficient\\
\rowcolor{LightGray}
$p_{\ell,k}, {\mathbf f}_{\mathrm{BB},\ell,k}$ & Transmit power and baseband precoder\\
$\mathbf{r}_{\ell}, \mathbf F_{\mathrm{RF},\ell}^{\varsigma}$ & RF chain activation vector and RF precoder\\
\rowcolor{LightGray}
$\mathbf{Q}_\ell, \boldsymbol{\epsilon}_{\ell}$ & DAC quantization loss matrix and noise vector
\\
$\mathbf{x}^\varsigma_{\ell,k},\mathbf{x}^\varsigma_{\ell,\mathrm{q}}$ & Effective HPC vector and RF-domain representation of DAC quantization noise vectors\\
\rowcolor{LightGray}
$\tilde{\mathbf{v}}_{\ell}, \mathbf{v}^\varsigma_\ell$ & Baseband and RF-domain transmit signal vector\\
$P_{\mathrm{cir},\ell},P_{\mathrm{tx},\ell},P_{\mathrm{tot}}$ & Circuit, transmit, and total power consumption\\
\rowcolor{LightGray}
$\eta_{\mathrm{PA}}$ & Power amplifier drain efficiency\\
$R_k,R_k^{\mathrm{(dm)}}$ & Achievable rate and traffic demand\\
\rowcolor{LightGray}
$R^\mathrm{(nr)}_{k},R^\mathrm{(nr)}_{\min}$ & Demand-normalized rate and its minimum value\\
\hline
\rowcolor{SectionGray}
\multicolumn{2}{l}{Algorithm notation} \\
\hline
$\breve{\mathcal K}_\ell$ & Candidate UE set\\
\rowcolor{LightGray}
$\mathcal W_\ell^\varsigma,\,\mathbf w_\ell$ & Joint assignment set and joint assignment vector\\
$\mathcal{C}_{\ell,i},\,\boldsymbol{\mathcal{C}}^{(n)}$ & The $i$-th feasible configuration of the $\ell$-th satellite and the $n$-th CE sample \\
\rowcolor{LightGray}
$N_s,\,N_{\mathrm{elite}}$ & Numbers of CE samples and elite samples\\
$\mathcal P, P_{\mathrm{com}}$ & Set of transmit powers and common power level\\
\rowcolor{LightGray}
$\beta_{\ell,k}$ & Transmit power coefficient of link $(\ell,k)$\\
\Xhline{0.8pt}
\end{tabular}
\end{table}

\begin{figure}[t]
    \centerline{\includegraphics[width=0.8\columnwidth]{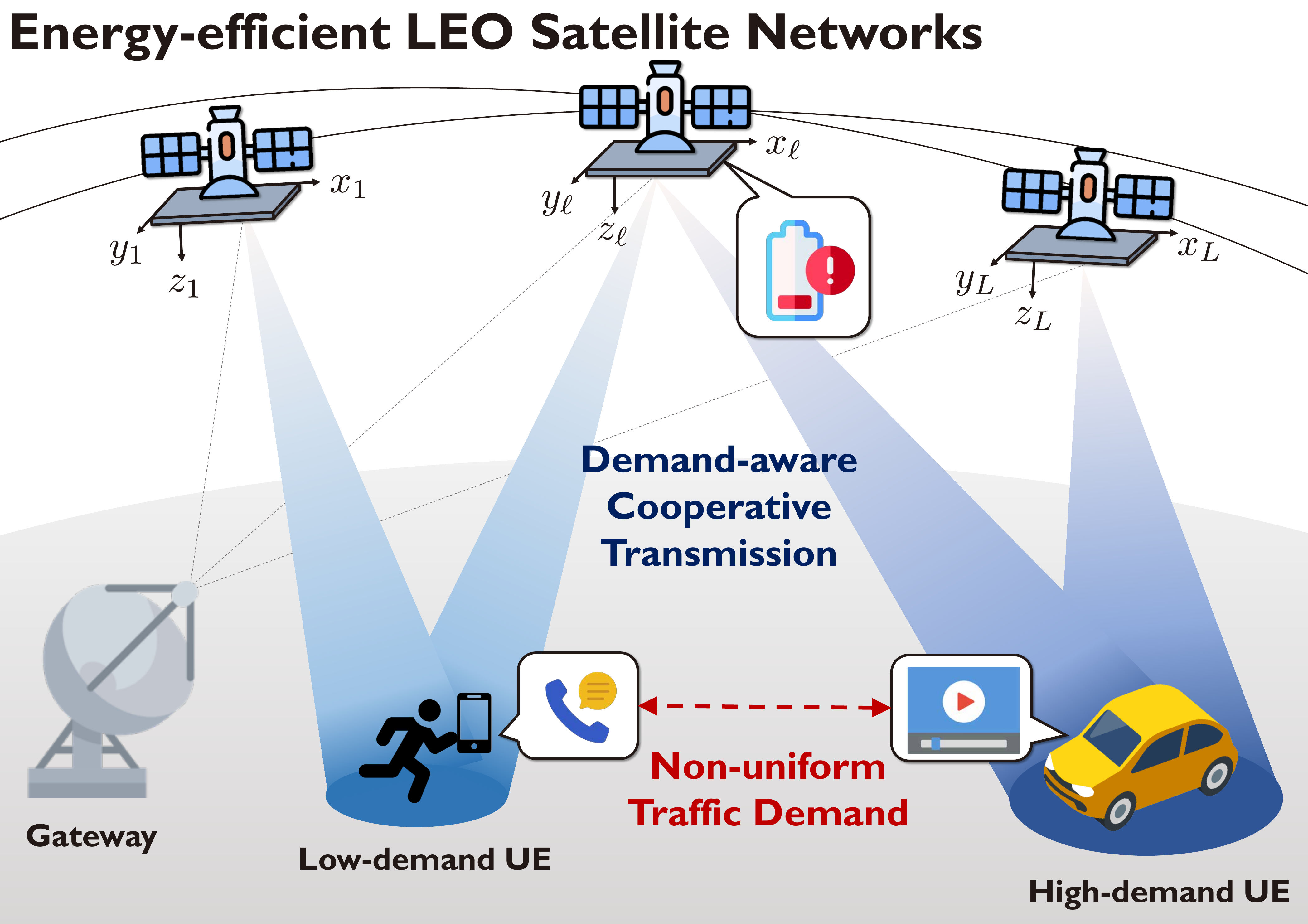
    }}
    \caption{Overall architecture of a cooperative LEO satellite network.} \label{fig:1}
\end{figure}

\vspace{-0.3cm}{\section{System Model} \label{section2}
Consider a network consisting of $L$ LEO satellites, $K$ single-antenna UEs with non-uniform traffic demands, and a gateway that coordinates cooperative transmissions from multiple satellites, as depicted in Fig.~\ref{fig:1}. For the $\ell$-th LEO satellite, a local coordinate system $(x_\ell, y_\ell, z_\ell)$ is used, where $z_\ell$ points toward the nadir, $x_\ell$ lies in the local orbital plane, and $y_\ell$ completes the right-handed coordinate system.
Each satellite is equipped with a uniform planar array (UPA) lying on the $x_{\ell}$-$y_{\ell}$ plane, with $N_\mathrm{x}$ and $N_\mathrm{y}$ elements along the $x_{\ell}$- and $y_{\ell}$-axes, respectively, so that the number of transmit antennas per satellite is $N_\mathrm{t} = N_\mathrm{x} N_\mathrm{y}$.
An HPC architecture with $N_{\mathrm{RF}} \leq N_{\mathrm{t}}$ RF chains is employed on each satellite. $\mathcal{K} = \{1,\ldots,K\}$, $\mathcal{K}_{\ell} \subseteq \mathcal{K}$, $\mathcal{L} = \{1,\ldots,L\}$, and $\mathcal{L}_k \subseteq \mathcal{L}$ denote the set of UEs, the set of UEs served by the $\ell$-th satellite, the set of LEO satellites, and the set of satellites jointly serving the $k$-th UE, respectively.
$\mathbf{p}^{\mathrm{(ue)}}_k\!\in\mathbb{R}^3$ and $\mathbf{p}^{\mathrm{(sat)}}_{\ell} \!\in\mathbb{R}^3$ are the position vectors of the $k$-th UE and the $\ell$-th LEO satellite, respectively, with both vectors expressed in an Earth-centered, Earth-fixed (ECEF) coordination system.}

\vspace{-0.2cm}\subsection{Channel Model}
To capture the LEO satellite channel characteristics, a ray-tracing channel model is adopted. Accordingly, the channel vector between the $\ell$-th LEO satellite and the $k$-th UE at the instant $t$ is given by
\begin{align}
    \mathbf{h}_{\ell,k}(t)\!=\!\!\!\!\!\! \sum^{M_{\ell,k}-1}_{m=0}\!\!\!\!\tilde \beta_{\ell,k,m}e^{j2\pi(\nu_{\ell,k,m}t\!-\!f_\mathrm{c}\! \tau_{\ell,k,m})}\mathbf{a}(\theta_{\ell,k,m},\!\phi_{\ell,k,m}).\label{eq:1}
\end{align}The channel comprises $M_{\ell,k}$ multipath components, where for the $m$-th path, $\tilde \beta_{\ell,k,m}$, $\nu_{\ell,k,m}$, and $\tau_{\ell,k,m}$ denote its complex gain, Doppler shift, and propagation delay, respectively, and $f_{\mathrm{c}}$ stands for the carrier frequency.
The array response vector $\mathbf{a}(\theta,\phi)$ corresponds to the off-nadir and azimuth AoD pair $(\theta,\phi)$, and is given by
\begin{align}
    \mathbf{a}(\theta,\phi)\triangleq \mathbf{a}_\mathrm{x}(\theta,\phi)\otimes\mathbf{a}_\mathrm{y}(\theta,\phi), \label{eq:2}
\end{align} 
where $\mathbf{a}_\mathrm{x}(\theta,\phi)\!\triangleq\! \bigl[1,e^{-j\pi \!\sin\!\theta \cos\!\phi},\cdots,e^{-j\pi(N_\mathrm{x}-1) \sin\theta \cos\phi}\bigr]^\top$ and $\mathbf{a}_\mathrm{y}(\theta,\phi)\!\triangleq\! \bigl[1,e^{-j\pi\!\sin\theta \sin\phi},\cdots,e^{-j\pi(N_\mathrm{y}-1) \sin\theta \sin\phi}\bigr]^\top$.
Owing to the high altitude of LEO satellites and the dominance of the line-of-sight (LoS) component, the DL channel exhibits the following three distinctive properties:
\begin{itemize}
    \item \textit{Angular spread}: Since the altitude of LEO satellites is much higher than that of the surrounding scatterers near each UE, the resulting angular spread is small\cite{3gpp38811}. Thus, the AoDs of all propagation paths can be approximated as identical, i.e., $(\theta_{\ell,k,m},\phi_{\ell,k,m}) \approx (\theta_{\ell,k},\phi_{\ell,k}),~\forall m$\cite{LiYou_1}.
    
    \item \textit{Doppler shift}: The Doppler shift of the $m$-th path is written as $\nu_{\ell,k,m} = \nu_{\ell,k,m}^{\mathrm{(sat)}} + \nu_{\ell,k,m}^{\mathrm{(ue)}}$, where $\nu_{\ell,k,m}^{\mathrm{(sat)}}$ arises from satellite mobility and $\nu_{\ell,k,m}^{\mathrm{(ue)}}$ from UE mobility. Due to the high altitude of the satellite, the Doppler shift caused by its motion varies negligibly across different paths, allowing the approximation $\nu_{\ell,k,m}^{\mathrm{(sat)}} \!\approx\! \nu_{\ell,k}^{(\mathrm{sat})},\forall m$\cite{Li_hybrid}.

    \item \textit{Delay spread}: The delay spread of LEO satellite channels is comparable to that of terrestrial channels, since most scattering takes place close to the ground\cite{3gpp38811}. The delay of each path can be expressed as $\tau_{\ell,k,m} = \tau_{\ell,k}^{\mathrm{(min)}} + \tau_{\ell,k,m}^{\mathrm{(res)}}$, where $\tau_{\ell,k}^{\mathrm{(min)}} \!\triangleq\! \min_m \tau_{\ell,k,m}$ and $\tau_{\ell,k,m}^{\mathrm{(res)}}$ denotes the residual delay of the $m$-th path.
\end{itemize}

By incorporating the above properties, the channel model in \eqref{eq:1} can be rewritten as
\begin{align}
    \mathbf{h}_{\ell,k}(t)
    = e^{j\Phi_{\ell,k}(t)}
      \,\breve{\beta}_{\ell,k}(t,f_\mathrm{c})\,
      \mathbf{a}_{\ell,k},
    \label{eq:3}
\end{align}
where
$
    \breve{\beta}_{\ell,k}(t,f_\mathrm{c})
    = \sum_{m=0}^{M_{\ell,k}-1} 
      \tilde \beta_{\ell,k,m}
      e^{j2\pi(\nu_{\ell,k,m}^{\mathrm{(ue)}} t 
      - f_\mathrm{c}\tau_{\ell,k,m}^{\mathrm{(res)}}),}$ represents the DL channel gain from the $\ell$-th satellite to the $k$-th UE, and
$
    \Phi_{\ell,k}(t)
    = 2\pi(\nu_{\ell,k}^{\mathrm{(sat)}}t
    - f_\mathrm{c}\tau_{\ell,k}^{\mathrm{(min)}})
$
represents the phase rotation determined by the satellite Doppler shift and the minimum propagation delay, both of which can typically be obtained from sCSI\footnote{The sCSI relies on geometry-based LoS parameters obtained from satellite ephemeris and UE positions, combined with long-term channel statistics. Owing to the high altitude and LoS-dominant nature of satellite links, the sCSI could remain valid over a certain time interval, even under rapid satellite motion.}\cite{kimcellfree}.
For notational simplicity, the array response vector is written as $\mathbf{a}_{\ell,k} \triangleq \mathbf{a}(\theta_{\ell,k},\varphi_{\ell,k})$.
Due to the LoS-dominant nature of LEO channels, $\breve{\beta}_{\ell,k}(t,f_\mathrm{c})$ is modeled as Rician fading. 
Using this model, \eqref{eq:3} can be rewritten as
\begin{align}
    \mathbf{h}_{\ell,k}(t)
    \!=\! e^{j\Phi_{\ell,k}(t) } \left(
        \sqrt{\frac{\gamma_{\ell,k}\kappa_{\ell,k}}{\kappa_{\ell,k}+1}}
        +
        \alpha_{\ell,k}
        \sqrt{\frac{\gamma_{\ell,k}}{\kappa_{\ell,k}+1}}
      \right)
      \mathbf{a}_{\ell,k}.
    \label{eq:4}
\end{align}
The large-scale fading coefficient, which accounts for free-space path loss, is expressed as $\gamma_{\ell,k} = G^{\mathrm{sat}} G^{\mathrm{ue}} \big(\frac{c}{4\pi f_{\mathrm{c}} d_{\ell,k}}\big)^{2}$. Here, $G^{\mathrm{sat}}$ and $G^{\mathrm{ue}}$ represent the directional transmit and receive antenna gains, respectively, while $d_{\ell,k} = \|\mathbf{p}^{(\mathrm{sat})}_{\ell} - \mathbf{p}^{(\mathrm{ue})}_{k}\|_2$ denotes the distance between the $\ell$-th LEO satellite and the $k$-th UE, and $c$ is the speed of light.  
The parameter $\kappa_{\ell,k}$ denotes the Rician K-factor, and $\alpha_{\ell,k}$ captures multipath effects, such as residual delay and Doppler shifts at the UE side, and is modeled as $\alpha_{\ell,k} \sim \mathcal{CN}(0,1)$.
This paper first assumes that the phase term $\Phi_{\ell,k}(t)$ in \eqref{eq:4}, arising from the satellite Doppler shift and the minimum propagation delay, is pre-compensated at the transmitter through the stream-wise baseband precoder by leveraging the predictable satellite motion. This pre-compensation assumption is commonly adopted in prior NTN studies, particularly in the cell-free NTN literature\cite{kimcellfree,ZhangISL,11328755}. Accordingly, the channel can be rewritten in the simplified form
$
    \mathbf{h}_{\ell,k}
    = \sqrt{\upsilon_{\ell,k}}
      \big(
        \sqrt{\kappa_{\ell,k}}
        +
        \alpha_{\ell,k}
      \big)\mathbf{a}_{\ell,k},
$
where $\upsilon_{\ell,k} = \gamma_{\ell,k} / (\kappa_{\ell,k}+1)$ is defined for notational simplicity. Later, the effect of imperfect pre-compensation is explicitly considered in the proposed algorithm in Section \ref{section5C}.

\begin{remark}
The proposed cooperative transmission model is aligned with the ongoing evolution toward coordinated, distributed, and multi-connectivity transmission scenarios \cite{3gpp38821,ITU2516}. However, in practical satellite networks, such cooperation requires not only accurate timing and phase synchronization, as reflected in a recent time-synchronization patent for LEO satellite clusters \cite{timesync}, but also reliable transport support via feeder links, backhaul, and inter-satellite link (ISL)-assisted paths. Under the assumed synchronization and transport support, the present work provides a forward-looking framework for evaluating the potential gains of multi-LEO cooperative transmission. Joint synchronization, transport-network control, and cooperative transmission design are left as important future work.
\end{remark}

\subsection{Signal Model}
In this paper, a cooperative DL transmission system that employs an HPC architecture is considered.
At the $\ell$-th LEO satellite, the baseband transmit signal vector is given by
\begin{align}
    \tilde{\mathbf{v}}_{\ell}
    = \mathbf{S}_{\mathcal{R}_{\ell}}
      \sum_{k\in\mathcal{K}_{\ell}} \tilde{\mathbf{f}}_{\mathrm{BB},\ell,k}\sqrt{p_{\ell,k}}\, s_k,
    \label{eq:5}
\end{align}where $s_k$ denotes the data symbol of the $k$-th UE with $\mathbb{E}\{s_k\}\!=\!0$ and $\mathbb{E}\{|s_k|^2\}\!=\!1$, and, for the $k$-th UE served by the $\ell$-th LEO satellite, $p_{\ell,k}$ is the transmit power and $\tilde {\mathbf{f}}_{\mathrm{BB},\ell,k}\in\mathbb{C}^{N_{\mathrm{RF}}\times 1}$ is the baseband precoder.
Let \(r_{\ell,n} \! \in \! \{0,1\}\) represent the status of the \(n\)-th RF chain of the \(\ell\)-th LEO satellite, where $r_{\ell,n} = 1$ indicates an active RF chain and $r_{\ell,n} = 0$ an inactive one. The corresponding RF chain activation vector is defined as \(\mathbf r_\ell \triangleq [r_{\ell,1},\ldots,r_{\ell,N_{\mathrm{RF}}}]^{\mathrm T}\).
Hence, the set of active RF chains for the \(\ell\)-th satellite is given by
\(
\mathcal{R}_{\ell} \!\triangleq\! \{\, n \!\in\! \{1,\dots,N_\mathrm{RF}\} \!\mid\! r_{\ell,n} \! = \! 1  \}.
\)
Let $\{n_{\ell,1} \!< \!\! \cdots \! < \!n_{\ell,|\mathcal{R}_{\ell}|}\}$ denote the ordered elements of $\mathcal{R}_{\ell}$.
The associated RF chain activation matrix is defined as  
$
\mathbf{S}_{\mathcal{R}_{\ell}}
\! = \!
[
\mathbf{e}_{n_{\ell,1}},\!\cdots\!,\mathbf{e}_{n_{\ell,|\mathcal{R}_\ell|}} 
]^{\!\top}
\!\in\!\mathbb{R}^{|\mathcal{R}_{\ell}|\times N_{\mathrm{RF}}},
$
where $\mathbf{e}_i\!\in\!\mathbb{R}^{N_{\mathrm{RF}}\times 1}$ is the unit vector whose $i$-th entry equals one.

The baseband transmit signal vector $\tilde{\mathbf{v}}_{\ell}$ is subsequently converted to the RF domain by the DACs. 
To model the effect of finite-resolution DACs, this work employs the AQNM\cite{Choi2022}, which represents the DAC output as a scaled version of the input with an uncorrelated distortion component.
The DAC output signal of the $\ell$-th LEO satellite is expressed as
\begin{align}
    Q(\tilde{\mathbf{v}}_{\ell}) = \mathbf{Q}_{\ell}\tilde{\mathbf{v}}_{\ell} + \boldsymbol{\epsilon}_{\ell}, \label{eq:6}
\end{align}
where $Q(\cdot)$ denotes an element-wise quantizer applied independently to the real and imaginary parts, and 
$
\mathbf{Q}_{\ell}
= \operatorname{diag}(q_{\ell,n_{\ell,1}},\ldots,q_{\ell,n_{\ell,|\mathcal{R}_\ell|}})\in\mathbb{C}^{|\mathcal{R}_\ell|\times|\mathcal{R}_\ell|}
$
represents the diagonal matrix of quantization loss, where each diagonal element $q_{\ell,n}$ denotes the quantization loss of the $n$-th RF chain.
Let \(b_{\mathrm{DAC}}\) denote the bit resolution of the DACs for all RF chains. For $b_{\mathrm{DAC}}\le 5$, each gain coefficient $q_\ell=q_{\ell,n}$ is taken from the lookup table in\cite{Li2015}, while for $b_{\mathrm{DAC}}>5$, the high-resolution approximation
$
q_{\ell}\! \approx \! 1 \!- \!\frac{\pi\sqrt{3}}{2} 2^{-2b_{\mathrm{DAC}}}
$
is used.
The vector $\boldsymbol{\epsilon}_{\ell}\sim\mathcal{CN}(\mathbf{0},\mathbf{R}_{\boldsymbol{\epsilon}_{\ell}})$ represents the additive quantization noise, whose covariance matrix is expressed as
$
    \mathbf{R}_{\boldsymbol{\epsilon}_{\ell}}
    \!=\! \mathbf{Q}_{\ell}\,\breve{\mathbf{Q}}_{\ell}\,
      \mathrm{diag} \big(\mathbb{E}\{\tilde{\mathbf{v}}_{\ell} \tilde{\mathbf{v}}_{\ell}^{\mathrm{H}}\}\big),
$
with $\breve{\mathbf{Q}}_{\ell}\!\triangleq\! \mathbf{I}_{|\mathcal{R}_\ell|}\! - \!\mathbf{Q}_{\ell}$ \cite{Choi2022}.

In the RF-domain, the RF precoder of the \(\ell\)-th LEO satellite, represented by \(\mathbf{F}^{\varsigma}_{\mathrm{RF},\ell}\), where \(\varsigma \!\in\! \{\mathrm{FC}, \mathrm{PC}\}\) indicates the chosen HPC architecture, is implemented using \(b_{\mathrm{PS}}\)-bit quantized PSs.
Accordingly, the RF-domain transmit signal vector at the $\ell$-th LEO satellite is defined as
\begin{align}
\mathbf{v}_{\ell}^{\varsigma}
&= \mathbf{F}_{\mathrm{RF},\ell}^{\varsigma}
Q\bigl(\tilde{\mathbf{v}_{\ell}}\bigr)
\label{eq:7}\\
&= \sum_{k\in\mathcal{K}_{\ell}}
\mathbf{x}_{\ell,k}^{\varsigma}
\sqrt{p_{\ell,k}}\,s_k
+ \mathbf{x}_{\ell,\mathrm{q}}^{\varsigma},
\label{eq:8}
\end{align}
where $\mathbf{x}^\varsigma_{\ell,k}
\triangleq \mathbf{F}^\varsigma_{\mathrm{RF},\ell}\mathbf{Q}_{\ell}\mathbf{S}_{\mathcal{R}_\ell}\tilde{\mathbf{f}}_{\mathrm{BB},\ell,k}$
represents the effective HPC vector for the $k$-th UE at the $\ell$-th LEO satellite, incorporating the baseband precoder, the RF chain activation matrix, the quantization loss matrix, and the RF precoder, and $\mathbf{x}_{\ell,\mathrm{q}}^{\varsigma}
\triangleq \mathbf{F}^\varsigma_{\mathrm{RF},\ell}\boldsymbol{\epsilon}_{\ell}$
denotes the RF-domain representation of the DAC quantization noise vector. Note that the quantization effects of the finite-resolution PSs forming the RF precoder and the DACs used for baseband processing are jointly incorporated in \eqref{eq:8}.

The baseband signal received at the $k$-th UE is then expressed as\cite{kimcellfree,ZhangISL}
\begin{align}
    y_{k} &= \sum_{ \ell\in\mathcal{L} } \mathbf h^{\mathrm{H}}_{\ell,k}  \mathbf{v}^\varsigma_{\ell} + z_{k} \label{eq:9} \\
    \nonumber &= \sum_{\ell \in\mathcal{L}_k} \mathbf h^{\mathrm{H}}_{\ell,k}  \mathbf{x}^\varsigma_{\ell,k} \sqrt{p
_{\ell,k}} s_k +\!\!\!\!\!\! \sum_{u \in\mathcal{K} \setminus \{ k\} } \! \sum_{\ell\in\mathcal{L}_u} \mathbf h^{\mathrm{H}}_{\ell,k}  \mathbf{x}^\varsigma_{\ell,u} \sqrt{p
_{\ell,u}} s_u \\
    &\quad + \sum_{\ell\in\mathcal{L}} \mathbf h^{\mathrm{H}}_{\ell,k} \mathbf{x}_{\ell,\mathrm{q}}^{\varsigma}   + z_{k}, \label{eq:10}
\end{align}
where $z_k \! \sim \! \mathcal{CN}(0,\sigma^2)$ denotes the noise observed at the $k$-th UE, and $\sigma^2$ is the corresponding noise variance.
The first term in \eqref{eq:10} represents the desired signal from satellites in $\mathcal{L}_k$, the second term represents the interference from satellites serving other UEs, and the third term accounts for the impact of DAC quantization noise as it propagates through the channels.

Given the long propagation delay and the short channel coherence time in LEO satellite communications, acquiring instantaneous channel state information (iCSI) is highly challenging and, even when obtained, it rapidly becomes outdated\cite{kimcellfree}. In cooperative transmission systems, coherent combining over large antenna arrays and multiple satellites induces channel hardening.
The aggregate effective channel coefficient at the $k$-th UE caused by the signal intended for the $u$-th UE is given as
\begin{align}
 \dot{g}_{k,u} 
    \triangleq
    \sum_{\ell\in\mathcal{L}_u}\!
        \mathbf{h}_{\ell,k}^{\mathrm{H}}
        \mathbf{x}^{\varsigma}_{\ell,u}
    \sqrt{p_{\ell,u}}, \label{eq:new11}
\end{align}which is concentrated around its statistical mean, thereby enabling reliable transceiver design based on sCSI.
To exploit this property, the received baseband signal is rewritten as
\begin{align}
    \nonumber
    y_k
    &= 
    \mathbb{E}\!\left\{ \dot{g}_{k,k} \right\} s_k
    \!+\! \Big( \dot{g}_{k,k} \!-\! \mathbb{E}\!\left\{ \dot{g}_{k,k} \right\} \Big) s_k
    \!+\!\!\!\! \sum_{u\in\mathcal{K}\setminus\{k\}} 
       \dot{g}_{k,u} s_u
    \\
    &\quad 
    + \sum_{\ell\in\mathcal{L}}
        \left( \ddot{\mathbf{g}}_{\ell,k} \right)^{\!\mathrm{H}}
        \boldsymbol{\epsilon}_{\ell}
    + z_k,
    \label{eq:11}
\end{align}
where
$
    \ddot{\mathbf{g}}_{\ell,k}
        \!=\! 
        (\mathbf{F}_{\mathrm{RF},\ell}^{\varsigma})^{\mathrm{H}}
        \mathbf{h}_{\ell,k}.
$

\vspace{-0.15cm}\subsection{Performance Metric}
\subsubsection{Achievable Rates}
By interpreting the first term in \eqref{eq:11} as the desired signal and grouping the remaining terms into an effective noise component, the resulting effective noise is zero-mean and uncorrelated with the transmitted symbol $s_k$. Under this model, the so-called \emph{use-and-then-forget} (UatF) bound, which is widely used in the massive MIMO literature\cite{9064545,8304782}, is adopted to derive a lower bound on the ergodic capacity based on sCSI, as shown in \eqref{eq:12} at the top of the next page.
\begin{figure*}[t]
\begin{align}
    R_k
    & \! = \!  
    \log_2\!\left(\!
    1 \!+ \!
    \frac{
        \big| \mathbb{E}\{ \dot{g}_{k,k} \} \big|^2
    }{
        \mathbb{V}\{ \dot{g}_{k,k} \}
        + \sum_{u \in \mathcal{K} \setminus \{k\} } \Big( \mathbb{V}\{ \dot{g}_{k,u}\} + \big|\mathbb{E}\{\dot{g}_{k,u}\}\big|^2 \Big)
        + \sum_{\ell\in\mathcal{L}}
          \mathrm{tr}\big(
            \mathbb{E}\{
              \ddot{\mathbf{g}}_{\ell,k}
              (\ddot{\mathbf{g}}_{\ell,k})^{\mathrm H}
              \} 
              \mathbf{R}_{\boldsymbol{\epsilon}_{\ell}}
          \big)
        + \sigma^{2}
    } \!
    \right) \label{eq:12}\\
    & \!= \! \log_2\!\left( \!1\! +\! 
        \frac{ 
        \Big| \! \sum\limits_{\ell\in\mathcal{L}_k}  \!\!
        \sqrt{ \upsilon_{\ell,k} \kappa_{\ell,k} p_{\ell,k}  } \mathbf{a}^\mathrm{H}_{\ell,k} \mathbf{x}^\varsigma_{\ell,k}  \Big|^2  
        }
        {  
        \sum\limits_{u \in \mathcal{K} } \! \sum\limits_{\ell\in\mathcal{L}_u} \!\!\!
        \upsilon_{\ell,k} p_{\ell,u}  \big| \mathbf{a}^\mathrm{H}_{\ell,k} \mathbf{x}^\varsigma_{\ell,u} \big |^2 
        \! + \!\!\!\!\!
        \sum\limits_{u \in \mathcal{K} \setminus \{k\} } \! \Big| \! \sum\limits_{\ell\in\mathcal{L}_u} \!\!\!
        \sqrt{ \upsilon_{\ell,k} \kappa_{\ell,k} p_{\ell,u}  } \mathbf{a}^\mathrm{H}_{\ell,k} \mathbf{x}^\varsigma_{\ell,u} \Big |^2 
        \!\! + \!\!
        \sum\limits_{\ell\in\mathcal{L}} \!
          \gamma_{\ell,k}\mathrm{tr}\big(
            (\mathbf{F}_{\mathrm{RF},\ell}^{\varsigma})^{\mathrm{H}}
        \mathbf{a}_{\ell,k} \mathbf{a}^\mathrm{H}_{\ell,k} \mathbf{F}_{\mathrm{RF},\ell}^{\varsigma}
              \mathbf{R}_{\boldsymbol{\epsilon}_{\ell}}
          \big)
        \! + \!
        \sigma^2 }\! \right)
    \label{eq:13}
\end{align}
\hrule
\vspace{-0.4cm}
\end{figure*} 
The mean and variance in \eqref{eq:12} can be expressed as
\begin{align}
    & \mathbb{E}\{ \dot{g}_{k,u}\} \!=\! \sum_{\ell\in\mathcal{L}_u} \!\!\!\sqrt{\upsilon_{\ell,k} \kappa_{\ell,k} p_{\ell,u} } \mathbf{a}^\mathrm{H}_{\ell,k} \mathbf{x}^{\varsigma}_{\ell,u},
    \label{eq:14}\\
    & \mathbb{V}\{ \dot{g}_{k,u} \} \!=\! \sum_{\ell\in\mathcal{L}_u} \upsilon_{\ell,k} p_{\ell,u} | \mathbf{a}^\mathrm{H}_{\ell,k} \mathbf{x}^{\varsigma}_{\ell,u}|^2, \label{eq:15}\\
    & \mathbb{E}\{ \ddot{\mathbf{g}}_{\ell,k}(\ddot{\mathbf{g}}_{\ell,k})^{\mathrm H}\} \!=\! (\mathbf{F}_{\mathrm{RF},\ell}^{\varsigma})^{\mathrm{H}}
        ( \gamma_{\ell,k} \mathbf{a}_{\ell,k} \mathbf{a}^\mathrm{H}_{\ell,k} ) \mathbf{F}_{\mathrm{RF},\ell}^{\varsigma}. \label{eq:16}
\end{align}
Substituting the above expressions into \eqref{eq:12} leads to \eqref{eq:13}.

\subsubsection{Power Consumption Model}
Next, a detailed model of power consumption is constructed by characterizing the hardware elements of the LEO satellite transmitter.
The DAC power consumption can be described by the empirical model
\(
P_{\mathrm{DAC}}(b_{\mathrm{DAC}})
= 1.5\times10^{-5}\,2^{b_{\mathrm{DAC}}}
  + 9\times10^{-12} f_{\mathrm{s}} b_{\mathrm{DAC}},
\)
indicating an exponential increase with the DAC resolution \(b_{\mathrm{DAC}}\) and a linear increase with the sampling rate $f_{\mathrm{s}}$\cite{8333733}.
The RF circuitry introduces a constant power component given by
\(
P_{\mathrm{RF}} = 2P_{\mathrm{LP}} + 2P_{\mathrm{M}} + P_{\mathrm{H}},
\)
where \(P_{\mathrm{LP}}\), \(P_{\mathrm{M}}\), and \(P_{\mathrm{H}}\) represent the power consumption of the low-pass filter, mixer, and hybrid with buffer, respectively \cite{Choi2022}.
To describe the power consumption of the PSs, let $N_{\mathrm{PS}}^{\varsigma}$ denote the number of PS elements per RF chain, which is equal to $N_\mathrm{t}$ for the FC architecture and to $N_\mathrm{sub} \!=\! N_\mathrm{t}/N_{\mathrm{RF}}$ for the PC architecture. Since each PS element draws $P_{\mathrm{PS}}(b_{\mathrm{PS}})$, the total power consumed by all PSs is $P_{\mathrm{PS}}(b_{\mathrm{PS}})N_{\mathrm{PS}}^{\varsigma}$.
Based on these components, the circuit power of the $\ell$-th LEO satellite can be expressed as\cite{Choi2022,8333733}

{ \vspace{-0.4cm} \begin{align}
P_{\mathrm{cir},\ell}
\! = \! P_{\mathrm{LO}}
\!+\!\! \sum_{n=1}^{N_{\mathrm{RF}}} \! r_{\ell,n} \!
\Big(
    2P_{\mathrm{DAC}}(b_{\mathrm{DAC}})
    \!+\! P_{\mathrm{RF}}
    \!+\! N_{\mathrm{PS}}^{\varsigma} P_{\mathrm{PS}}(b_{\mathrm{PS}})
\!\Big),
\label{eq:17}
\end{align}
}\noindent
where $P_{\mathrm{LO}}$ is the local oscillator power.
Concurrently, the transmit power at the $\ell$-th LEO satellite is expressed as

\vspace{-0.5cm}
\begin{align}
P_{\mathrm{tx},\ell}
= \mathrm{tr}\!\left(\mathbb{E}\!\left\{
\mathbf{v}^{\varsigma}_{\ell}(\mathbf{v}^{\varsigma}_{\ell})^{\mathrm{H}}
\right\}\right). \label{eq:18}
\end{align}
The total power consumption of the LEO satellite network is thereby defined as

\vspace{-0.5cm}
\begin{align}
P_{\mathrm{tot}}
= \sum_{\ell\in\mathcal{L}} P_{\mathrm{cir},\ell}
  + \sum_{\ell\in \mathcal{L}} 
  \eta^{-1}_{\mathrm{PA}} P_{\mathrm{tx},\ell},
\label{eq:19}
\end{align}
where $\eta_{\mathrm{PA}}$ is the power amplifier drain efficiency.

\section{Problem Formulation} \label{section3}
\subsection{Demand-Aware Energy Efficiency Maximization}

This paper aims to simultaneously capture both power consumption and non-uniform traffic demands in LEO satellite networks. Existing demand-aware objectives, such as wMMF and rate-matching, account for UE requirements\cite{ratematching,WMMF}, but the corresponding optimization problems typically depend on manually specified power constraints to account for power consumption. In contrast, conventional EE uses a unified objective that trades off SE and power consumption\cite{Choi2022,Li_hybrid,Liu2024}, but does not explicitly model traffic demand, and adding quality-of-service (QoS) constraints can make the problem infeasible in some cases. 
To address this, the demand-normalized rate and its minimum value are first defined as

\vspace{-0.4cm}
\begin{align}
    &R_k^{(\mathrm{nr})} \triangleq R_k /R_k^{(\mathrm{dm})},\quad \forall k \in \mathcal{K}, \label{eq:new20}\\
    &R^{(\mathrm{nr})}_{\min} \triangleq \min_{k \in \mathcal{K}} R_k^{(\mathrm{nr})}, \label{eq:new21}
\end{align}
where $R_k^{(\mathrm{dm})}\!$ denotes the traffic demand of the $\!k$-th UE.
Using these definitions, the demand-aware EE is proposed as

\vspace{-0.4cm}
\begin{align}
    f(\mathcal{V}_0)
    = \frac{  R_{\min}^\mathrm{(nr)} (\mathcal{V}_0)}{P_\mathrm{tot}(\mathcal{V}_0)},
    \label{eq:20}
\end{align}where $\mathcal{V}_0 \! \triangleq \!
\{
\mathbf{F}^{\varsigma}_{\mathrm{RF},\ell},
\mathcal{K}_\ell,
\mathbf{r}_{\ell},
\tilde{\mathbf{f}}_{\mathrm{BB},\ell,k}, 
p_{\ell,k}
 |  
k\in\mathcal{K},\ell\in\mathcal{L}
\}$
denotes the design variable set.
By employing inverse-demand weighting together with a max–min formulation\cite{WMMF}, this objective promotes rate allocations that scale proportionally to UE demands, while at the same time taking the total network energy consumption into account. 
With these definitions, the optimization problem is formulated as

{\vspace{0cm}\color{black}
\begin{maxi!}
    { \scriptstyle \mathcal{V}_0} 
    { \!\!\!\! f(\mathcal{V}_0) } 
    {\label{eq:21}} 
    {\hspace{-0.1cm}\mathscr{P}_0\!:} 
    \addConstraint{ \!\!\!\! \mathrm{tr}\big(\mathbb{E}\{\mathbf{v}^{\varsigma}_\ell(\mathcal{V}_0) (\mathbf{v}^{\varsigma}_\ell(\mathcal{V}_0))^{\mathrm H}\}\big)\le P_{\mathrm{max}},\, \forall \ell\in\mathcal{L}}{}{} \label{eq:21b}
    \addConstraint{ \!\!\!\!\| \mathbf{x}^\varsigma_{\ell,k} (\mathcal{V}_0) \|^2=1,\, \forall k\in\mathcal{K}_\ell,\ell\in\mathcal{L}}{}{}  \label{eq:21d}
    \addConstraint{ \!\!\!\! p_{\ell,k} \geq 0,\, \forall k\in\mathcal{K}_\ell, \ell\in\mathcal{L}}{}{} \label{eq:21c}\addConstraint{\!\!\!\!\mathbf{F}^{\varsigma}_{\mathrm{RF},\ell}\in\mathcal{F}^\varsigma_{\mathrm{RF},\ell
    },\, \forall l\in\mathcal{L}}{}{}  \label{eq:21e}
    \addConstraint{ \!\!\!\![\mathbf{r}_\ell]_n \! \in \!\{0,1\}, n\!=\! 1,\!\ldots\!,N_{\mathrm{RF}},\forall \ell\!\in\!\mathcal{L}}{}{} \label{eq:21f}
    \addConstraint{\!\!\!\!|\mathcal{K}_\ell|\le \|\mathbf{r}_\ell \|_0 ,\, \forall \ell\in\mathcal{L}}{}{}  \label{eq:21h}
    \addConstraint{\!\!\!\!\cup_{\ell\in\mathcal{L}} \mathcal{K}_\ell = \mathcal{K}.}{}{}  \label{eq:21i}
\end{maxi!}
}\noindent
Constraint~\eqref{eq:21b} imposes the per-satellite maximum transmit power $P_{\max}$, and \eqref{eq:21d} ensures that the effective HPC vectors are normalized, and \eqref{eq:21c} guarantees that the allocated power remains nonnegative. Constraint \eqref{eq:21e} restricts the RF precoder to the feasible FC/PC set $\mathcal{F}^\varsigma_{\mathrm{RF},\ell}$, which consists of matrices whose entries are drawn from the $b_{\mathrm{PS}}$-bit discrete PS (DPS) codebook and satisfy the unit-modulus constraint, i.e., $|[\mathbf{F}^{\varsigma}_{\mathrm{RF},\ell}]_{m,n}| \!=\! 1$. Constraint \eqref{eq:21f} ensures that the RF chain activation variables are binary and that the number of active RF chains at each satellite does not exceed $N_{\mathrm{RF}}$. Constraint~\eqref{eq:21h} links UE association to RF chain activation by ensuring that the number of UEs served by the $\ell$-th LEO satellite does not exceed that of its active RF chains. Finally, constraint \eqref{eq:21i} enforces that each UE be associated with at least one satellite.

\vspace{-0.15cm} \section{Algorithm Design} \label{section4}
\begin{figure}[t]
    \centerline{\includegraphics[width=0.75\columnwidth]{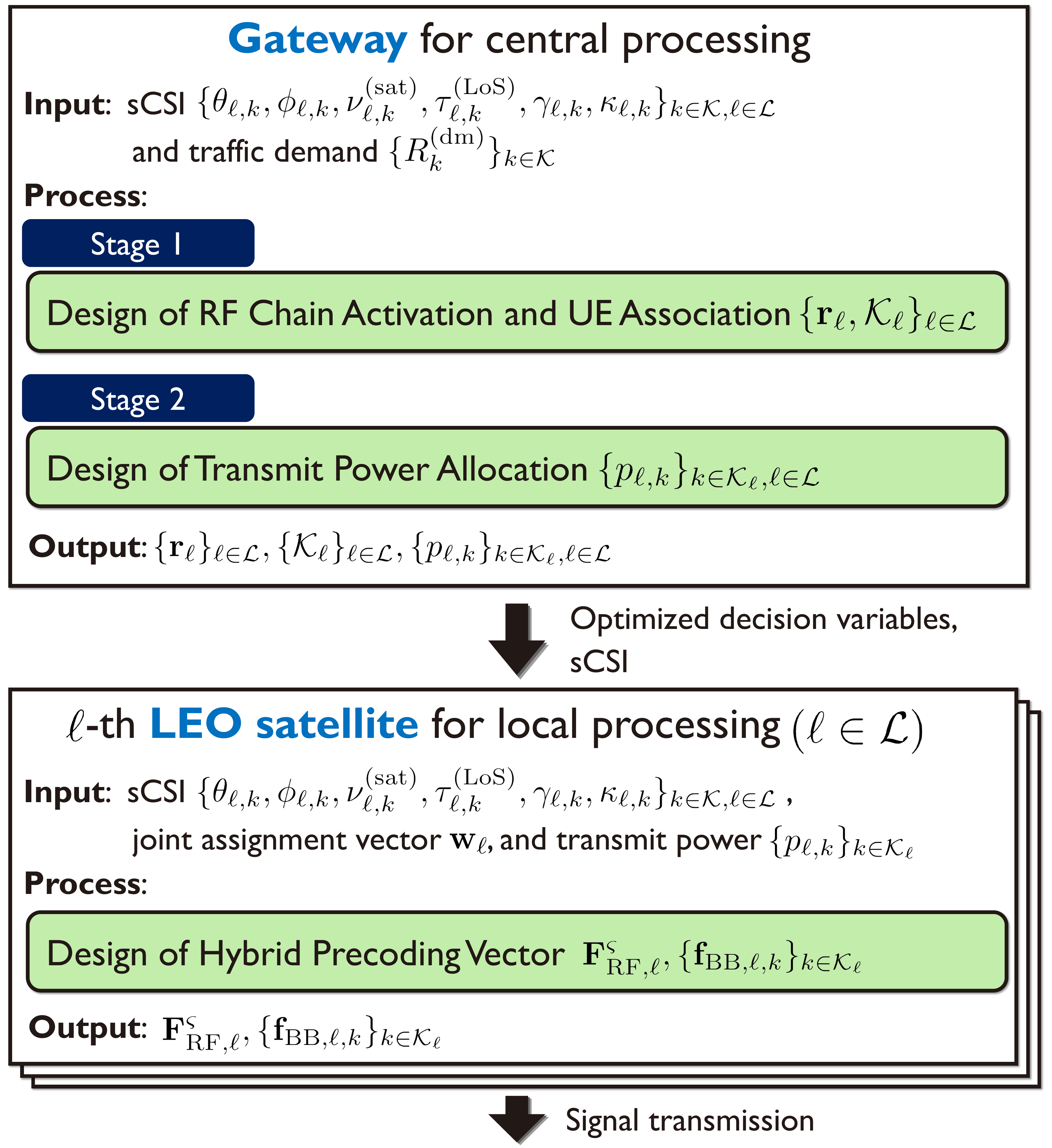}}
    \caption{Overall transmission procedure and design variable decomposition.} \label{fig:2}
    \vspace{-0.7em}
\end{figure}

The resulting nonconvex optimization, which couples continuous and discrete design variables, leads to MINLP whose global solution is computationally prohibitive. Naive cooperative transmission architectures would also require the CPU to jointly optimize all variables with high computational complexity and exchange them with the satellites, which is impractical under limited wireless backhaul. To this end, the proposed algorithm employs a divide-and-conquer approach to obtain a high-quality solution. After collecting sCSI and traffic demands from the UEs, the CPU jointly determines the RF chain activation, UE association, and transmit power allocation, forwards these low-dimensional decisions to the satellites, and then each satellite locally computes its own linear precoder and transmits the data signals.\footnote{Because each satellite constructs its own precoders locally, the gateway does not need to forward either the RF or the baseband precoder. This avoids signaling overhead that scales with the full precoder dimension and instead makes the overhead grow only linearly with key system parameters (i.e., numbers of satellites, RF chains, and UEs), when other parameters are fixed.} The overall transmission procedure is depicted in Fig.~\ref{fig:2}.

\subsection{Design of the HPC Vector} 
Due to the single-AoD characteristic of satellite channels \cite{3gpp38811}, the DL channel from the $\ell$-th LEO satellite to the $k$-th UE is nearly aligned with its associated array response vector. Hence, it is reasonable to design the RF precoder along the array response vector to maximize the array gain \cite{ZhangISL}. Moreover, since this vector is uniquely determined by the satellite–UE geometry and can be directly obtained from positional information, it naturally fits the sCSI-based design. However, the RF precoder is implemented using $b_{\mathrm{PS}}$-bit DPSs, which determine the set of admissible phase values. Accordingly, for the $\ell$-th satellite serving the $k$-th UE on its $n$-th active RF chain, the RF precoder $\mathbf{f}^\varsigma_{\mathrm{RF},\ell,n} \!\in\! \mathbb{C}^{N_{\mathrm t}\times1}$ is specified element-wise as \cite{kyeongsoo}
\begin{align}
    [\mathbf{f}^\varsigma_{\mathrm{RF},\ell,n}]_i 
    = \exp\Bigl(j\Bigl(\tfrac{2\pi \dot{m}_{\ell,k}^\star }{2^{b_{\mathrm{PS}}}}
      + \tfrac{\pi}{2^{b_{\mathrm{PS}}}}\Bigr)\Bigr), \label{eq:22}
\end{align}
where the optimal phase index $\dot{m}_{\ell,k}^\star$ is given by
\begin{align}
   \dot{m}_{\ell,k}^\star 
    = \mathop{\mathrm{argmin}}\limits_{\dot{m}_{\ell,k}=0,\ldots,2^{b_{\mathrm{PS}}}-1}
      \Bigl|\angle [\mathbf{a}_{\ell,k}]_i 
      - \Bigl(\tfrac{2\pi \dot{m}_{\ell,k} }{2^{b_{\mathrm{PS}}}}
      + \tfrac{\pi}{2^{b_{\mathrm{PS}}}}\Bigr)\Bigr|. \label{eq:23}
\end{align}
Here, $i$ denotes the antenna-element index. For the FC architecture, the $n$-th RF chain is connected to all $N_\mathrm{t}$ antennas and $i\!\in\!\{1,\ldots,N_\mathrm{t}\}$, whereas for the PC architecture, it is connected to a disjoint subarray of size $N_{\mathrm{sub}}$, where $i\!\in\!\{(n-1)N_{\mathrm{sub}}\!+\!1,\ldots,nN_{\mathrm{sub}}\}$. 
The remaining entries of $\mathbf{f}^\mathrm{PC}_{\mathrm{RF},\ell,n}$ are set to zero. 
This approach naturally enforces the unit-modulus constraint, fulfilling \eqref{eq:21e} without the need for sophisticated optimization procedures, and concentrates the transmit energy toward the UEs without optimization.

However, due to the wide footprint of satellite beams, energy focusing alone cannot sufficiently control inter-UE interference. Therefore, the baseband precoder is designed to suppress the remaining inter-UE interference. To incorporate the effects of the RF precoder, DAC quantization, and the single-AoD channel structure, the effective channel of the \(\ell\)-th satellite is defined as
\(
\mathbf{H}^{\mathrm{eff}}_\ell \! \triangleq  \! \mathbf{A}^{\mathrm{H}}_\ell \, \mathbf{F}^{\varsigma}_{\mathrm{RF},\ell} \, \mathbf{Q}_\ell,
\)
where
\(
\mathbf{A}_\ell \triangleq [\mathbf{a}_{\ell,1}, \ldots, \mathbf{a}_{\ell,K}].
\)
Accordingly, the reduced-dimensional baseband precoder for the $\ell$-th LEO satellite and the $k$-th UE,
$
\mathbf{f}_{\mathrm{BB},\ell,k}
\! \triangleq \! \mathbf{S}_{\mathcal{R}_\ell}\tilde{\mathbf{f}}_{\mathrm{BB},\ell,k}
\! \in \! \mathbb{C}^{|\mathcal{R}_\ell|\times 1},
$
is specified as a normalized\footnote{ The effective hybrid precoding vector does not introduce additional power scaling. Consequently, the transmit power allocated to each stream is determined solely by the power variable \(p_{\ell,k}\), thereby keeping the baseband precoder normalization consistent with the transmit power optimization.} zero-forcing (ZF)\footnote{ While maximum ratio transmission (MRT) and regularized ZF (RZF) are also possible choices, ZF is selected as a representative baseband precoder because it provides a practical balance between inter-UE interference suppression and computational complexity.} vector\cite{6457363,kimcellfree} 
\begin{align}
\mathbf{f}_{\mathrm{BB},\ell,k} =
\frac{
\left(\mathbf{H}^{\mathrm{eff}}_\ell\right)^{\mathrm{H}}
\bigl(\mathbf{H}^{\mathrm{eff}}_\ell(\mathbf{H}^{\mathrm{eff}}_\ell)^{\mathrm{H}}\bigr)^{-1}
\mathbf{e}_k
}{
\bigl\|
\mathbf{F}^{\varsigma}_{\mathrm{RF},\ell}\mathbf{Q}_\ell
\left(\mathbf{H}^{\mathrm{eff}}_\ell\right)^{\mathrm{H}}
\bigl(\mathbf{H}^{\mathrm{eff}}_\ell(\mathbf{H}^{\mathrm{eff}}_\ell)^{\mathrm{H}}\bigr)^{-1}
\mathbf{e}_k
\bigr\|
}, \label{eq:24}
\end{align}
which mitigates inter-UE interference.

The proposed precoder is efficient but introduces a new issue regarding UE–RF chain association. As indicated in \eqref{eq:22} and \eqref{eq:23}, designing the RF precoder for the $n$-th RF chain entails deciding which UE will be served by the RF chain. This association can be expressed via the mapping  
$\Pi_\ell : \mathcal{R}_\ell \to \mathcal{K}_\ell$, where $\Pi_\ell(n)$ represents the UE associated with the $n$-th RF chain. The requirement $\Pi_\ell(\mathcal{R}_\ell) = \mathcal{K}_\ell$ guarantees that all UEs are served, thereby satisfying constraint \eqref{eq:21i}. The mapping $\Pi_\ell$ is considered as part of the RF chain activation and UE association problem in Section~\ref{section4B}.

\vspace{-0.1em} \subsection{Design of RF Chain Activation, UE Association, and Transmit Power Allocation} \label{section4B}
After designing the HPC precoders as in the previous subsection, the optimization problem in \eqref{eq:21} is rewritten as\begin{maxi!}
    { \scriptstyle \mathcal{V}_1} 
    { f(\mathcal{V}_1)} 
    {\label{eq:25}} 
    {\mathscr{P}_1:} 
    \addConstraint{ \nonumber\eqref{eq:21b},\,\eqref{eq:21c},\,\eqref{eq:21f},\,\eqref{eq:21h}  }{}{}
    \addConstraint{ \Pi_\ell(\mathcal{R}_\ell(\mathbf{r}_\ell)) = \mathcal{K}_\ell,\, \forall \ell\in\mathcal{L} }{}{} \label{eq:pi25}
\end{maxi!}using the variable set 
\(
\mathcal{V}_1 \!\triangleq\! 
\{
\Pi_\ell, \mathbf{r}_\ell, \mathcal{K}_\ell,  p_{\ell,k}
 |  k\in\mathcal{K}, \ell\in\mathcal{L},
\Pi_\ell\!:\!\mathcal{R}_\ell\!\to\!\mathcal{K}_\ell
\}
\).
Let $\mathcal{V}_0(\mathcal{V}_1)$ denote the complete variable set obtained by recovering the precoders via \eqref{eq:22}--\eqref{eq:24}. Define $f(\mathcal{V}_1) \triangleq f(\mathcal{V}_0(\mathcal{V}_1))$, and analogously for $R_{\min}^\mathrm{(nr)}$ and $P_{\mathrm{tot}}$. 
Note that the constraint in \eqref{eq:21h} behaves differently for the different architectures due to the RF precoder structure in \eqref{eq:22}. In the FC architectures, multiple RF chains assigned to the same UE yield identical RF precoders; activating extra RF chains only provides power gain without changing the RF precoder subspace, so $|\mathcal{K}_\ell| \! = \! \|\mathbf{r}_\ell\|_0$ is naturally enforced. In the PC architectures, each RF chain drives a distinct subarray, so additional RF chains for a UE at each satellite provide extra array gain, allowing $|\mathcal{K}_\ell| \!\le\! \|\mathbf{r}_\ell\|_0$. Consequently, $\mathscr{P}_0$ has architecture-dependent constraints and thus differs between the FC and PC architectures, as will become clear in the sequel.

Since the resulting problems remain MINLP, a two-stage approach is adopted to obtain a high-quality solution, in which the RF chain activation and the UE association are jointly optimized in the first stage, and the transmit power is optimized in the second stage.

\subsubsection{\textbf{Stage 1: Design of RF Chain Activation and UE Association}}
The search space for the mapping $\{\Pi_\ell\}_{\ell=1}^L$ induced by RF chain activation and UE association grows as $(\sum_{n=0}^{N_{\mathrm{RF}}}\binom{K}{n}\frac{N_{\mathrm{RF}}!}{(N_{\mathrm{RF}}-n)!})^{L}$ in the FC architecture, and as $(K+1)^{L N_{\mathrm{RF}}}$ in the PC architecture, which makes the exhaustive search impractical even for moderate values of $L$. 
To cope with this problem, the search space of each satellite is restricted to a candidate UE set, denoted by $\breve{\mathcal{K}}_\ell \subseteq \mathcal{K}$, which is formed by taking into account the characteristics of the satellite channel. Because the satellite operates at a high altitude, channel quality is mainly determined by the distance, and UEs that are closely spaced are hard to distinguish in the angular domain.
Let $N_\ell^{(\mathrm{v})}$ denote the number of UEs visible to the $\ell$-th satellite, $k_{\ell,(i)}$ denote the $i$-th nearest visible UE, and $\vartheta_{\ell,k,u}$ denote the angular separation between the $k$-th and $u$-th UEs as seen from the $\ell$-th satellite.
Starting with $\breve{\mathcal K}_{\ell}^{(1)}\!=\!\{k_{\ell,(1)}\}$, the candidate UE set is recursively built as
\begin{align}
\breve{\mathcal K}_{\ell}^{(i)}
\!=\!
\begin{cases}
\breve{\mathcal K}_{\ell}^{(i-1)}\!\cup\!\{k_{\ell,(i)}\},
&
\!\!\!\!\! \min\limits_{u\in\breve{\mathcal K}_{\ell}^{(i-1)}}
\vartheta_{\ell,k_{\ell,(i)},u}
\!>\!
\vartheta_{\mathrm{th}},\\
\breve{\mathcal K}_{\ell}^{(i-1)},
&\!\!\!\!\!
\text{otherwise},
\end{cases}
\end{align}
for $i\!=\!2,
\!\ldots\!,N_{\ell}^{\mathrm{(v)}}$, where $\vartheta_{\mathrm{th}}$ is the angular-separation threshold, yielding the candidate UE set $\breve{\mathcal K}_{\ell} \triangleq \breve{\mathcal K}_{\ell}^{(N_{\ell}^{\mathrm{(v)}})}$.

To efficiently investigate the refined search space, the CE method\cite{CEbj,CEbook} is employed, which reformulates the original deterministic optimization problem as a stochastic one. In this method, the algorithm repeatedly refines the sampling distributions by maximizing the objective, causing them to gradually converge to the distributions associated with the optimal solution. Because the optimal distribution is not known in advance, it is inferred by performing repeated sampling. For clarity, a detailed procedure is presented below.

\indent $\vcenter{\hbox{\tiny$\bullet$}}~$\textbf{Step 1 (Definition and Initialization)}: 
To define the sampling distributions for each realization of RF chain activation and UE association, the joint assignment set is first defined as
{\vspace{-0.5em}
\begin{align}
\nonumber \mathcal{W}_\ell^\varsigma = \Big\{ \mathbf{w}_\ell &= [ w_{\ell,1}, \ldots, w_{\ell,|\breve{\mathcal{K}}_\ell|}]^\top \in \mathbb{Z}_+^{|\breve{\mathcal{K}}_\ell| \times 1} \\
&\cdots \Big|\ 0 \le w_{\ell,i} \le w_{\max}^{\varsigma},\ \sum_{i=1}^{|\breve{\mathcal{K}}_\ell|} w_{\ell,i} \le N_{\mathrm{RF}} \Big\}, \label{eq:26}
\end{align}
}\noindent
where \(w_{\max}^{\varsigma} \!=\! 1\) for the FC architectures and \(w_{\max}^{\varsigma} \!=\! N_{\mathrm{RF}}\) for the PC architectures.
Let $\{k_{\ell,1} \!<\! \cdots \!<\! k_{\ell,|\breve{\mathcal{K}}_{\ell}|}\}$ denote the ordered elements of $\breve{\mathcal{K}}_{\ell}$. Here, $w_{\ell,1}$ represents the number of RF chains that the $\ell$-th satellite allocates to the $k_{\ell,1}$-th UE in $\breve{\mathcal{K}}_{\ell}$. In other words, each joint assignment vector $\mathbf{w}_\ell \!\in\! \mathcal{W}_\ell^\varsigma$ specifies the number of RF chains allocated to each UE in $\breve{\mathcal{K}}_\ell$. Rather than deciding which specific RF chain is assigned to each UE, the joint set only specifies how many RF chains are utilized by each UE. 
Focusing on this multiplicity is well motivated since the satellite--UE distance is much larger than the array aperture, so the transmission performance is mainly affected by how many RF chains are allocated to a UE rather than which specific subarrays are selected. 
Accordingly, \(\mathcal{W}_\ell^\varsigma\) provides a convenient way to implement \(\Pi_\ell \!:\! \mathcal{R}_\ell \! \to \! \mathcal{K}_\ell\) under a naive sequential assignment strategy.\footnote{Starting from the UE with the smallest index (e.g., UE\,1), if UE\,1 is allocated two RF chains, the RF chains with indices 1 and 2 are assigned to UE\,1. Then, the process moves on to the next UE, and so on. In this way, the mapping \(\Pi_\ell\) determines which RF chain is assigned to which UE.}

To specify which configuration in the joint assignment set $\mathcal{W}_\ell^\varsigma$ is adopted at each satellite, let $\mathcal{C}_{\ell,i}$ denote an $i$-th feasible configuration set for the $\ell$-th satellite, where $i=1,\ldots,|\mathcal{W}_\ell^\varsigma|.$
Each joint assignment vector $\mathbf{w}_\ell \in \mathcal{W}^\varsigma_\ell$ is matched with a single $\mathcal{C}_{\ell,n}$, and this set is defined in order to represent which configuration the satellite belongs to.
For the $\ell$-th satellite, a single configuration $\ell \in \mathcal{C}_{\ell,i}$ is selected, which uniquely specifies both the number of active RF chains and the associated UEs.
All configurations selected in this way automatically satisfy constraints \eqref{eq:21f} and \eqref{eq:21h}.
The probability mass function (PMF) for configuration selection of the $\ell$-th satellite is defined through the vector $\boldsymbol{\rho}_\ell \in \mathbb{R}^{|\mathcal{W}_\ell^\varsigma|\times 1}$ as
\begin{align}
    [\boldsymbol{\rho}_\ell]_i 
    = \rho_{\ell,i}
    \triangleq 
    \Pr\{\ell\in\mathcal{C}_{\ell,i}\}, \label{eq:27}
\end{align}
subject to the normalization constraint $\sum_{i=1}^{|\mathcal{W}_\ell^\varsigma|} \rho_{\ell,i} = 1.$
Let $\boldsymbol{\mathcal{C}}\!=\! \{ \mathcal{C}_{1,1},\ldots,\mathcal{C}_{L,|\mathcal{W}_L^\varsigma|} \}$
denote a configuration selection across the $L$ satellites.
For notational simplicity, define \(\boldsymbol{\rho} \triangleq \{\boldsymbol{\rho}_\ell\}_{\ell\in\mathcal{L}}\); given \(\boldsymbol{\rho}\), the probability of observing 
$\boldsymbol{\mathcal{C}}$ is \cite{CEbj}

\vspace{-0.4cm}
\begin{align}
\mathsf{P}(\boldsymbol{\mathcal{C}}; \boldsymbol{\rho} )
=
\prod_{\ell=1}^{L}
\sum_{i=1}^{|\mathcal{W}_\ell^\varsigma|}
\rho_{\ell,i}\,
\mathbbm{1}_{\{ \ell \in \mathcal{C}_{\ell,i}\}}, \label{eq:28}
\end{align}
where $\mathbbm{1}_{\{\cdot\}}$ denotes the indicator function.
Initially, $\boldsymbol{\rho}$ is set to a uniform distribution.

\indent $\vcenter{\hbox{\tiny$\bullet$}}~$\textbf{Step 2 (Sampling and Evaluation):}
Based on the distribution in \eqref{eq:28}, $N_{\mathrm{s}}$ samples of configuration selections are generated across the $L$ satellites and denoted by
$
\{\boldsymbol{\mathcal{C}}^{(n)}\}_{n=1}^{N_{\mathrm{s}}},
$
where the $n$-th sample configuration (hereafter referred to as the CE sample) is given by
$
\boldsymbol{\mathcal{C}}^{(n)}
=
\{\mathcal{C}^{(n)}_{1,1},\ldots,\mathcal{C}^{(n)}_{L,|\mathcal{W}_L^\varsigma|}\}.
$
The set of transmit powers is defined as 
\begin{align}
    \mathcal{P} \!\triangleq\! \{ p_{\ell,k} \!\mid\! k\in\mathcal{K}_\ell,\ell\in\mathcal{L} \}, \label{eq:powervar}
\end{align}
and, for the $n$-th CE sample, the objective is evaluated as

\vspace{-0.3cm}
\begin{align}
\!\!\!\!f_{\mathrm{CE}}(\boldsymbol{\mathcal{C}}^{(n)}, \! \mathcal{P})
\! \triangleq \! 
\left\{
\begin{array}{lr}
-\infty, & \!\!\!\! \exists\, k\in\mathcal{K}\ \text{s.t.}\ \mathcal{L}_k=\varnothing,\\
\frac{
R_{\min}^\mathrm{(nr)}(\boldsymbol{\mathcal{C}}^{(n)}, \mathcal{P})
}{
P_{\mathrm{tot}}(\boldsymbol{\mathcal{C}}^{(n)} \mathcal{P})
},
& \!\!\!\! \text{otherwise}.
\end{array}
\right.\label{eq:29}
\end{align}
Given $(\boldsymbol{\mathcal{C}}^{(n)},\!\mathcal{P})$, the variable set $\mathcal{V}_1$ is uniquely determined; hence, $\!f_\mathrm{CE}\!$ can be written as functions of $(\boldsymbol{\mathcal{C}}^{(n)},\!\mathcal{P})$.
To satisfy \eqref{eq:pi25}, $-\infty$ is included. However, for all other CE samples, evaluating \eqref{eq:29} requires assigning suitable transmit power. As the sample size increases, optimizing the transmit power for each CE sample becomes more impractical, whereas using a constant transmit power overlooks the selection-dependent effects.

To address this challenge, the transmit power for each UE is initially allocated in proportion to its traffic demand, i.e., $p_{\ell,k} = (R_k^{\mathrm{(dm)}} /  R_{\ell}^{\mathrm{(sum)}}) P_{\mathrm{com}}$, where $R_{\ell}^{\mathrm{(sum)}} \triangleq \sum_{u \in \mathcal{K}_\ell} R_u^{\mathrm{(dm)}}$. Subsequently, the common power level $P_{\mathrm{com}}$, which scales all transmit powers, is optimized. This approach reduces the transmit power allocation problem to finding a single scalar and enables efficient evaluation for each CE sample.
Let $\zeta_{\ell} \triangleq \sum_{k\in\mathcal{K}_\ell }\!\beta_{\ell,k} \!\cdot \! R_k^\mathrm{(dm)}/R_\ell^\mathrm{(sum)}$, where $\beta_{\ell,k}$ is the coefficient of $p_{\ell,k}$ associated with DAC quantization noise and signal transmission.\footnote{The transmit power is affected by the transmission of the information signal and DAC quantization noise. Their combined contribution to the transmit power is represented by a coefficient derived in Appendix \ref{appendix:A}.}
For a feasible CE sample $\boldsymbol{\mathcal C}^{(n)}$, the common power level $P_{\mathrm{com}}$ is obtained by solving\cite{Choi2022}
\begin{maxi!}
  {\scriptstyle P_{\mathrm{com}}}
  { {f}  (P_{\mathrm{com}}) }
  {\label{eq:30}}
  {\mathscr{P}_2(\boldsymbol{\mathcal C}^{(n)}):}
  \addConstraint{0 \le P_{\mathrm{com}} \le P_{\mathrm{eff}} \label{eq:31b} ,}
\end{maxi!}
where \(P_\mathrm{eff} \! \triangleq \! P_{\max}/\max_\ell \zeta_{\ell}\) ensures that the common power level remains within the transmit power limit of the strongest satellite, in accordance with constraint~\eqref{eq:21b}. With $\boldsymbol{\mathcal{C}}^{(n)}$ fixed, its dependence on $f$ is omitted in $\mathscr{P}_2$.

Since $f(P_{\mathrm{com}})$ is unimodal in $[0, P_{\mathrm{eff}}]$, the unique maximizer of $\mathscr{P}_2(\boldsymbol{\mathcal C}^{(n)})$ can be efficiently found through a simple one-dimensional search, with unimodality proved in Appendix \ref{appendix:B}.
To implement the one-dimensional search efficiently, the differentiation of $f(P_{\mathrm{com}})$ with respect to $P_{\mathrm{com}}$ is used,

\vspace{-0.3cm}
\begin{align}
  f'( P_{\mathrm{com}})  \triangleq  \frac{N^\mathrm{(sg)} ( P_{\mathrm{com}})}{ P_\mathrm{tot} (P_{\mathrm{com}})^2}. \label{eq:31}
\end{align}
For notational simplicity, the common argument \(P_{\mathrm{com}}\) is omitted. Then,
$
  N^\mathrm{(sg)}
  \! \triangleq \!
  (R_{\min}^{\mathrm{(nr)}})' P_{\mathrm{tot}}
\! -\! 
  R_{\min}^{\mathrm{(nr)}} P_{\mathrm{tot}}',
$
where all terms are evaluated at \(P_{\mathrm{com}}\). 
Since the denominator of \(f'\) is positive, its sign is entirely determined by that of \(N^\mathrm{(sg)}\), which is strictly decreasing on \([0, P_\mathrm{eff}]\) and satisfies \(N^\mathrm{(sg)}(0) > 0\) (see Appendix \ref{appendix:B}).
Consequently, the boundary value at $P_\mathrm{eff}$ is first checked: if $ N^\mathrm{(sg)} ( P_{\mathrm{eff}} ) \!\ge\! 0$, then $ f$ is strictly increasing in $[0, P_\mathrm{eff}]$, and the optimal solution is $P_{\mathrm{com}}^\star \! = \! P_{\mathrm{eff}}$, without any optimization. Otherwise, there exists a unique $P_{\mathrm{com}}^\star \!\in\! [0,P_{\mathrm{eff}})$ such that $N^\mathrm{(sg)}(P_{\mathrm{com}}^\star) \!=\! 0$, and this point can be rapidly found by bisection search \cite{WCNCwooseok}.

To evaluate $N^\mathrm{(sg)}$, $(R_{\min}^\mathrm{(nr)})'$ and $P'_\mathrm{tot}$ need to be calculated.
The function $R_{\min}^\mathrm{(nr)}(P_{\mathrm{com}})$ is concave but, in general, only piecewise differentiable (see Appendix~\ref{appendix:B}).
Hence, the sub-differential of $R_{\min}^\mathrm{(nr)}(P_{\mathrm{com}})$ is introduced,
\(
  \partial R_{\min}^\mathrm{(nr)}(P_{\mathrm{com}})
  \! \triangleq \! \{ \xi \!\in\! \mathbb{R} \mid
  R_{\min}^\mathrm{(nr)}(x) \!\le\! R_{\min}^\mathrm{(nr)} (P_{\mathrm{com}})
  \!+\! \xi(x - P_{\mathrm{com}}), \forall x \!\in\! [0,P_{\mathrm{eff}}] \},
\)
and the extremal subgradients are given by $\max \partial R_{\min}^{\mathrm{(nr)}}(P_{\mathrm{com}})$ and $\min \partial R_{\min}^{\mathrm{(nr)}}(P_{\mathrm{com}})$.
And since 
\(
P_\mathrm{tot}(P_{\mathrm{com}}) \!= \!\sum_{\ell\in\mathcal{L}} P_{\mathrm{cir},\ell}
  \!+\! \eta_{\mathrm{PA}}^{-1} \sum_{\ell\in\mathcal{L}}\zeta_{\ell} P_{\mathrm{com}}
\)
can be rewritten as a function of \(P_\mathrm{com}\), we have
\(
P'_\mathrm{tot}(P_{\mathrm{com}})\! \triangleq \!\eta_{\mathrm{PA}}^{-1} \sum_{\ell\in\mathcal{L}} \zeta_{\ell}.
\)
Using these quantities, one can compute the minimum value $N^\mathrm{(sg)}_{\min}(P_{\mathrm{com}})$ and the maximum value $N^\mathrm{(sg)}_{\max}(P_{\mathrm{com}})$ of $N^\mathrm{(sg)}(P_{\mathrm{com}})$.

Using these values, a bisection search over $P_{\mathrm{com}} \!\!\in\!  [0, P_{\mathrm{eff}})$ is performed as detailed in Algorithm~\ref{alg:1}.
\begin{algorithm}[!t]
    \caption{Bisection search for $P_{\mathrm{com}}$ in $\mathscr{P}_2$}
    \label{alg:1}
    \begin{algorithmic}
    \State \textbf{Input:} $\boldsymbol{\mathcal C}$, $P_{\mathrm{eff}}$ and $\varepsilon_{\mathrm{B}}$
    \State \textbf{Initialize:} $P_{\mathrm{L}} \gets 0$, $P_{\mathrm{U}} \gets P_{\mathrm{eff}}$, $\dot{t} \gets 0$
    \If{$N_{\max}(P_{\mathrm{eff}}) \ge 0,N_{\min}(P_{\mathrm{eff}}) \ge 0$}
      \State \textbf{return} $P_{\mathrm{com}}^\star \gets P_{\mathrm{eff}}$
    \Else
        \While{(1)}
            \State Set $P_{\mathrm{mid}} \gets (P_{\mathrm{L}} + P_{\mathrm{U}})/2$
            \If{$|N^\mathrm{(sg)}_{\min}(P_{\mathrm{mid}})| \!\le\! \varepsilon_{\mathrm{B}}$ \textbf{or} $|N^\mathrm{(sg)}_{\max}(P_{\mathrm{mid}})| \!\le\! \varepsilon_{\mathrm{B}}$}
                \State \textbf{break} 
            \ElsIf{$N^\mathrm{(sg)}_{\min}(P_{\mathrm{mid}}) > 0$}
                \State Update $P_{\mathrm{L}} \gets P_{\mathrm{mid}}$ 
            \ElsIf{$N^\mathrm{(sg)}_{\max}(P_{\mathrm{mid}}) < 0$}
                \State Update $P_{\mathrm{U}} \gets P_{\mathrm{mid}}$ 
            \ElsIf{$N^\mathrm{(sg)}_{\min}(P_{\mathrm{mid}}) \!\le\! 0$ \textbf{and} $N^\mathrm{(sg)}_{\max}(P_{\mathrm{mid}})\!\ge\! 0$ }
                \State \textbf{break} 
            \EndIf
            \State $\dot{t} \gets \dot{t} + 1$
        \EndWhile
    \EndIf
    \State \textbf{return:} $P_{\mathrm{com}}^\star \gets P_{\mathrm{mid}}$
  \end{algorithmic}
\end{algorithm}
After evaluating the samples, they are sorted in descending order of the objective value as
$
 f_{\mathrm{CE}}(\boldsymbol{\mathcal{C}}^{(1)},\! P^{(1)}_{\mathrm{com}})
 \!\ge\!\!
 \cdots
 \!\ge\!\!
 f_{\mathrm{CE}}(\boldsymbol{\mathcal{C}}^{(N_{\mathrm{s}})},\! P^{(N_{\mathrm{s}})}_{\mathrm{com}}),
$
where \(P_{\mathrm{com}}^{(n)}\!\!\) is the optimal common power level for the \(n\)-th CE sample.

\indent $\vcenter{\hbox{\tiny$\bullet$}}~$\textbf{Step 3 (Estimation and Update)}:
In this step, the size of the elite set is determined by \( N_{\mathrm{elite}} = \lceil \lambda N_{\mathrm{s}} \rceil \), where \(\lambda \in (0,1]\) denotes the elite ratio, and this size is then used to define the elite threshold \( f_{\mathrm{th}} = f_{\mathrm{CE}}\big( \boldsymbol{\mathcal{C}}^{(N_{\mathrm{elite}})}, P^{(N_{\mathrm{elite}})}_{\mathrm{com}} \big). \)
Once the elite threshold has been determined, the elite set is defined as
\(
\mathcal{E
} \!\triangleq\! 
\{
1,\ldots,N_{\mathrm{elite}}
\}.
\)
Using the elite samples, a sampling distribution that favors these top-performing realizations can be constructed.
This leads to the following CE-based optimization over the set of PMF vectors $\boldsymbol{\rho}$, as in \cite{CEbook}:
\begin{maxi!}
    {\scriptstyle \boldsymbol{\rho}}
    {
        \frac{1}{N_{\mathrm{s}}}
        \sum_{n=1}^{N_{\mathrm{s}}}
        \mathbbm{1}_{\{n\in\mathcal{E}\}}
        \ln \mathsf{P}(\boldsymbol{\mathcal{C}}^{(n)};\boldsymbol{\rho})
    }
    {\label{eq:36}}
    {\mathscr{P}_3(\{ \boldsymbol{\mathcal{C}}^{(n)} \}_{n=1}^{N_\mathrm{s}} )\!:\!}
    \addConstraint{
        \sum_{i=1}^{|\mathcal{W}_\ell^\varsigma|} \rho_{\ell,i} = 1,
    }{}
    {\qquad \forall\,\ell\in\mathcal{L}.} \label{eq:36b}
\end{maxi!}
This problem can be solved directly by applying the Lagrangian formulation under the probability constraint, where the Lagrangian is defined as
\begin{align}
    \nonumber \mathscr{L}(\boldsymbol{\rho},\{\mu_\ell\}_{\ell=1}^L) 
\triangleq & 
\frac{1}{N_{\mathrm{s}}}
\sum_{n=1}^{N_{\mathrm{s}}}
\mathbbm{1}_{\{n\in\mathcal{E}\}}
\ln \mathsf{P}(\boldsymbol{\mathcal{C}}^{(n)};\boldsymbol{\rho})
\\&+ 
\sum_{\ell=1}^{L}
\mu_\ell
\Big(1-\sum_{i=1}^{|\mathcal{W}_\ell^\varsigma|}\rho_{\ell,i}\Big), \label{eq:37}
\end{align}
with the Lagrangian multiplier $\mu_\ell$ for $\ell\!=\!1,\!\ldots\!,L$.
Applying the stationarity condition
\(
\frac{\partial \mathscr{L} (\boldsymbol{\rho},\{\mu_\ell\}_{\ell=1}^L)}{\partial \rho_{\ell,i}} \!=\! 0
\)
leads to
\(
    \rho_{\ell,i}
\! = \!
\frac{
\sum_{n=1}^{N_{\mathrm{s}}}\!
\mathbbm{1}_{\{n\in\mathcal{E}\}}\!
\mathbbm{1}_{\{\ell\in\mathcal{C}_{\ell,i}\}}
}{N_{\mathrm{s}}\,\mu_\ell}.\!
\)
Using the feasibility constraint \eqref{eq:36b},
the optimal multipliers and PMFs are obtained as
\begin{align} 
&\mu_\ell^\star
    = 
    \frac{1}{N_{\mathrm{s}}}
    \sum_{n=1}^{N_{\mathrm{s}}}
    \mathbbm{1}_{\{n\in\mathcal{E}\}}=\frac{N_{\mathrm{elite
    }} }{N_{\mathrm{s}} },  \\
& \rho_{\ell,i}^\star \! =\!
    \frac{
        \sum_{n=1}^{N_{\mathrm{s}}}\!\!
        \mathbbm{1}_{\{n\in\mathcal{E}\}}\!
        \mathbbm{1}_{\{\ell\in\mathcal{C}_{\ell,i}\}}
    }{
        \sum_{n=1}^{N_{\mathrm{s}}}
        \mathbbm{1}_{\{n\in\mathcal{E}\}}
    }\!\!=\!\!\frac{\sum_{n=1}^{N_{\mathrm{s}}}\!\!
        \mathbbm{1}_{\{n\in\mathcal{E}\}} \!
        \mathbbm{1}_{\{\ell\in\mathcal{C}_{\ell,i}\}} }{N_{\mathrm{elite}} },\label{eq:39}
\end{align} 
for $\ell\in\mathcal{L}$ and $i=1,\ldots,|\mathcal{W}_\ell^\varsigma|$.
From \eqref{eq:39}, it is intuitive that the optimal distribution gives each configuration a probability equal to the proportion of elite samples in which it appears.
With the smoothing update
{\vspace{0cm} \begin{align}
    \boldsymbol{\rho}_\ell^{(\ddot{t}\,)}
    = \digamma \, \boldsymbol{\rho}_\ell^{\star}
    + (1-\digamma) \, \boldsymbol{\rho}_\ell^{(\ddot{t}-1)},\,\forall \ell \in \mathcal{L}, \vspace{-0.4cm} \label{eq:40} 
\end{align}}\noindent
where $\ddot{t}$ is the iteration index and $\digamma \!\in\! (0,1)$ is a smoothing factor, the distribution is gradually steered toward the elite selections.
Because the CE procedure uses only a few hyperparameters, the update is mostly self-tuning and needs little human intervention.
After convergence, the resulting configuration for the $\ell$-th LEO satellite is obtained as
\begin{align}
i_\ell^{\star}
=
\mathop{\mathrm{argmax}}\limits_{i_\ell =1,\ldots,|\mathcal{W}_\ell^\varsigma|} \rho_{\ell,i},    \label{eq:41}
\end{align}
which automatically determines both the activate RF chains and the associated UEs.
The detailed procedure of the modified CE method is summarized in Algorithm \ref{alg:2}.
\begin{algorithm}[t]
  \caption{Modified CE-Method for $\boldsymbol{\rho}$ in $\mathscr{P}_3$}
  \label{alg:2}
  \begin{algorithmic}
    \State \textbf{Input:} $\{ \mathcal{W}_\ell^\varsigma \}_{\ell\in\mathcal{L}}$, $N_{\mathrm{s}}$, $N_{\mathrm{elite}}$, $\digamma$, and $T_{\mathrm{CE}}$
    \State \textbf{Initialize:} $\ddot{t} = 0$ and $\rho_{\ell,k}^{(\ddot{t}\,)} = 1/|\mathcal{W}_\ell^\varsigma|$
    \Repeat
      \State Generate configuration sets $\{ \boldsymbol{\mathcal{C}}^{(n)} \}_{n=1}^{N_{\mathrm{s}}}$ from \eqref{eq:28}
      \State Calculate $\{ f_\mathrm{CE}(\boldsymbol{\mathcal{C}}^{(n)},P^{(n)}_\mathrm{com}) \}_{n=1}^{N_{\mathrm{s}}}$ using Algorithm \ref{alg:1}
      \State Define elite set $\mathcal{E}$ using $N_{\mathrm{elite}}$ 
      \State Obtain $\rho_{\ell,k}^{\star}$ using \eqref{eq:39}  
      \State Update $\boldsymbol{\rho}^{(\ddot{t}+1)}$ using \eqref{eq:40}
      \State Set $\ddot{t} \leftarrow \ddot{t} + 1$
    \Until{convergence or $\ddot{t} = T_{\mathrm{CE}}$}
    \State \textbf{return:} $\!\!\{\mathbf{w}^\star_\ell\}_{\ell\in\mathcal{L}}\!$ using \eqref{eq:41} and sequential assignment
  \end{algorithmic}
\end{algorithm}

\subsubsection{\textbf{Stage 2: Design of Transmit Power Allocation}}
With the obtained RF chain activation and UE association, the remaining task is to optimize the transmit power to maximize the demand-aware EE. 
Let $\{\ell_{k,1}\!<\!\!\cdots\!<\!\!\ell_{k,|\mathcal{L}_k|}\}$ denote the ordered elements of $\mathcal{L}_k$, and introduce the $k$-th UE-centric square-root transmit power vector as
$
    \mathbf{p}_k
    \!=\!
    [
        \sqrt{p_{\ell_{k,\!1},k}},
        \!\ldots,\!
        \sqrt{p_{\ell_{k,\!|\mathcal{L}_k|},k}}
    ]^\top \!\! \in \! \mathbb{R}^{|\mathcal{L}_k|\times1}.
$
Then, \eqref{eq:13} is rewritten as
{\vspace{0cm}\begin{align}
    & R_k \!=\! \log_2\!\big(1+\Upsilon_k\big), \nonumber\\
    & \Upsilon_k \!\!=\!
    \frac{
        (\mathbf{g}^\top_k \mathbf{p}_k )^2
    }{
        \sum\limits_{u\in\mathcal{K}} \!\!
            \|\mathbf{G}^{(\mathrm{v})}_{k,u}\mathbf{p}_u\|^2
        \!+\!\!\!\!\!\!\!\!
        \sum\limits_{u\in\mathcal{K}\setminus\{k\}}\!\!\!\!\!\!\!\!
            \|\mathbf{G}^{(\mathrm{e})}_{k,u}\mathbf{p}_u\|^2
        \!+\!\!\!
        \sum\limits_{u\in\mathcal{K}} \!\!
            \|\mathbf{G}^{(\mathrm{q})}_{k,u}\mathbf{p}_u\|^2
        \!+\!
        \sigma^2
    } ,
\label{eq:42}
\end{align}} \noindent
where $\mathbf{g}_k \in \mathbb{R}^{|\mathcal{L}_k|\times 1}$,
$\mathbf{G}^{(\mathrm{v})}_{k,u} \in \mathbb{R}^{|\mathcal{L}_u|\times|\mathcal{L}_u|}$,
$\mathbf{G}^{(\mathrm{e})}_{k,u} \in \mathbb{R}^{2\times|\mathcal{L}_u|}$, and
$\mathbf{G}^{(\mathrm{q})}_{k,u} \in \mathbb{R}^{|\mathcal{L}_u|\times|\mathcal{L}_u|}$.
Their nonzero entries are defined as
{\vspace{0cm}\begin{align}
    &[\mathbf{g}_k]_i
    \! = \!\!
    \sqrt{\upsilon_{\ell_{k,i},k}\kappa_{\ell_{k,i},k} }
    \mathbf{a}^\mathrm{H}_{\ell_{k,i},k}
    \mathbf{x}^{\varsigma}_{\ell_{k,i},k},
    \, i=1,\ldots,|\mathcal{L}_k| , \label{eq:43}
\\
    &[\mathbf{G}^{(\mathrm{v})}_{k,u}]_{i,i}
    \! = \!\! 
    \sqrt{
            \upsilon_{\ell_{u,i},k}
    }
    \vert \mathbf{a}^\mathrm{H}_{\ell_{u,i},k}
    \mathbf{x}^{\varsigma}_{\ell_{u,i},u} \vert,\, i=1,\ldots,|\mathcal{L}_u| , \label{eq:44}
\\
    \nonumber &[\mathbf{G}^{(\mathrm{e})}_{k,u}]_{:,i}
    \!=\! [
        \Re\!\left\{
            \Lambda_{k,u,i}
        \right\}
        ,
        \Im\!\left\{
            \Lambda_{k,u,i}
        \right\}
    ]^\top,  \\ 
    &\Lambda_{k,u,i}
    \! = \!\!
    \sqrt{
            \upsilon_{\ell_{u,i},k}\kappa_{\ell_{u,i},k}
    }
    \mathbf{a}^\mathrm{H}_{\ell_{u,i},k}
    \mathbf{x}^{\varsigma}_{\ell_{u,i},k},\, i=1,\ldots,|\mathcal{L}_u| , \label{eq:45}
\\
    & [\mathbf{G}^{(\mathrm{q})}_{k,u}]_{i,i}
    \!=\!\!
    \sum_{n=1}^{|\mathcal{R}_\ell|} \!
    \gamma_{\ell_{u,i},k}
    [\tilde{\mathbf{Q}}^{\varsigma}_{\ell_{u,i},k}]_{n,n}
    \big|\!
        [\tilde{\mathbf{f}}_{\mathrm{BB},\ell_{u,i},k}]_n
    \!\big|^2\!,i\!\!=\!1,\!\ldots\!,|\mathcal{L}_u|, \label{eq:46}
\end{align}}\noindent
with 
$
\tilde{\mathbf{Q}}^{\varsigma}_{\ell,k}
    =
    (\mathbf{F}^\varsigma_{\mathrm{RF},\ell})^\mathrm{H}
    \mathbf{a}_{\ell,k}\mathbf{a}^{\mathrm{H}}_{\ell,k}
    \mathbf{F}^{\varsigma}_{\mathrm{RF},\ell}
    \mathbf{Q}_{\ell} \breve{\mathbf{Q}}_{\ell},\forall \ell\!\in\!\mathcal{L},\,k\!\in\!\mathcal{K}.
$ 
Consequently, the problem of power allocation becomes
{\begin{maxi!}
    {\scriptstyle \mathcal{P} }
    {
    \frac{
        \min_{k\in\mathcal{K}}
        R_k(\mathcal{P})/R^{\mathrm{(dm)}}_k
    }{
        \eta_{\mathrm{PA}}^{-1}
        \! \sum\limits_{\ell\in \mathcal{L}} \!
        \sum\limits_{k\in\mathcal{K}_\ell}\beta_{\ell,k}p_{\ell,k}\!+\!
        P_{\mathrm{cir}}
    }}
    {\label{eq:47}}
    {\mathscr{P}_4: }
    \addConstraint{
        \sum_{k\in\mathcal{K}_\ell} \beta_{\ell,k} p_{\ell,k} \le P_{\max},\forall\,\ell\in\mathcal{L} 
    }{}{} \label{eq:47b}
    \addConstraint{
        p_{\ell,k}\ge 0,\, \forall\,k\in\mathcal{K}_\ell ,\ell\in\mathcal{L},
    }{}{} \label{eq:47c}
\end{maxi!}\noindent
where $R_k$ can be expressed as a function of $\mathcal{P}$, the set of transmit powers $\{p_{\ell,k}\}_{k\in\mathcal{K}_\ell,\ell\in\mathcal{L}}$ as defined in \eqref{eq:powervar}, with all other design variables fixed. For compactness, we define $P_{\mathrm{cir}} \triangleq \sum_{\ell\in\mathcal{L}} P_{\mathrm{cir},\ell}$, and let $\beta_{\ell,k}$ denote the previously defined coefficient associated with the transmit power.
Since the fractional programming (FP) problem $\mathscr{P}_4$ is difficult to solve directly, a slack variable $\nu$ is introduced, and the resulting problem is transformed into an equivalent subtractive form using the Dinkelbach method \cite{dinkelbach1967nonlinear}, as follows:
\begin{maxi!}
    {\scriptstyle \mathcal{P},\nu}
    {
        \nu
        \!-\!
        \chi^\star
        \Big(
            \eta_{\mathrm{PA}}^{-1}\!
            \sum_{\ell\in\mathcal{L}}\!
            \sum_{k\in\mathcal{K}_\ell}\!\beta_{\ell,k}p_{\ell,k}\!
            +\!
            P_{\mathrm{cir}}       
        \Big)
    }
    {\label{eq:48}}
    { \dot{\mathscr{P}}_4(\chi^\star):}
    \addConstraint{\nonumber \eqref{eq:47b},\eqref{eq:47c} }{}{}
    \addConstraint{
        \nu \le  R_k(\mathcal{P})/R^{\mathrm{(dm)}}_k,\,\forall\,k\in\mathcal{K}.
    }{}{} \label{eq:48b}
\end{maxi!}For the optimal objective value $\chi^\star$ of \eqref{eq:47}, the associated parametric problem $\dot{\mathscr{P}}_4(\chi^\star)$ attains the optimal value zero. Since $\chi^\star$ is unknown, $\chi$ is updated iteratively and the parametric subproblem is repeatedly solved until the objective function is sufficiently close to zero. At iteration $\check{t}$, the update is given by
\begin{align}
    \chi^{(\check{t}\,)}
    =
    \frac{
        \min_{k} R_k(\mathcal{P}^{{(\check{t}-1)}} )/R^{\mathrm{(dm)}}_k
    }{
        \eta_{\mathrm{PA}}^{-1}
        \sum_{\ell\in\mathcal{L}}
        \sum_{k\in\mathcal{K}_\ell} \beta_{\ell,k} p^{{(\check{t}-1)}}_{\ell,k}
        +
        P_{\mathrm{cir}}
    }, \label{eq:49}
\end{align}
where $\mathcal{P}^{(\check{t}-1)} \triangleq \{\, p^{(\check{t}-1)}_{\ell,k} \! \mid \! \ell\!\in\!\mathcal{L},\ k\!\in\!\mathcal{K}_\ell \,\}$
denotes the power allocation obtained at the $(\check{t}-1)$-th Dinkelbach iteration.
However, this problem remains non-convex due to the constraint in \eqref{eq:48b}. 
By applying the quadratic transform to convexify the formulation \cite{Quadratictransform}, an equivalent parametric optimization problem is obtained, where the parameters $\psi^{(\hat{t}\,)}_k$ are iteratively updated and slack variables $\omega_k$, $\forall k$, are introduced. The resulting optimization problem is given as
\begin{subequations}\label{eq:51eq}
\begin{align}
{\!\!\!\! \ddot{\mathscr{P}}}_4 ( \chi^{(\check{t}\,)}\!,&\psi^{(\hat{t}\,)} )\!: \underset{\mathcal{V}_4}{\max}\ 
\nu
        \!-\!\!
        \chi^{(\check{t})}
        \Big(
            \eta_{\mathrm{PA}}^{-1}
            \sum_{\ell\in\mathcal{L}}\!
            \sum_{k\in\mathcal{K}_\ell}\!\!\beta_{\ell,k}p_{\ell,k}
            \!+\!\!
            P_{\mathrm{cir}}            
        \Big), \label{eq:51a}\\
\mathrm{s.t.}\quad
&\nonumber \eqref{eq:47b},\ \eqref{eq:47c}, \\
&\nu \le \log_2(1+\omega_k)/R_k^{\mathrm{(dm)}},\ \forall k\in\mathcal{K}, \\
&\omega_k \le 2\psi_k^{(\hat{t})}\mathbf{g}_k^\top\mathbf{p}_k
-(\psi_k^{(\hat{t})})^2D_k(\mathcal{P}),\ \forall k\in\mathcal{K}. 
\end{align}
\end{subequations}
where the design variable set is
$\mathcal{V}_4 \triangleq \{ \mathcal{P}, \{\omega_k\}_{k\in\mathcal{K}}, \nu \}$.
Note that the denominator of the signal-to-interference-plus-noise ratio (SINR) in \eqref{eq:42}
is introduced for notational simplicity as
\(
D_k(\mathcal{P}) =
\sum_{u\in\mathcal{K}} \|{\mathbf{G}}^{(\mathrm{v})}_{k,u} {\mathbf{p}}_u\|^2
+
\sum_{u\in\mathcal{K}\setminus \{k\} }\|{\mathbf{G}}^{(\mathrm{e})}_{k,u}{\mathbf{p}}_u\|^2
+
\sum_{u\in\mathcal{K}}\|{\mathbf{G}}^{(\mathrm{q})}_{k,u}{\mathbf{p}}_u\|^2+
\sigma^2.
\)
Using the transmit power from the previous quadratic transform iteration, the auxiliary variable at iteration $\hat{t}$ is given by
\begin{align}
    \psi_k^{(\hat{t})}
    =
    \big| \mathbf{g}^\top_k \, \mathbf{p}^{(\hat{t}-1)}_k \big| / 
    D_k\big(\mathcal{P}^{(\hat{t}-1)}\big),
    \label{eq:51}
\end{align}

\noindent where $\mathcal{P}^{(\hat{t})}$ denotes the power allocation obtained at the $\hat{t}$-th quadratic transform iteration.
Since ${\ddot{\mathscr{P}}}_4 ( \chi^{(\check{t}\,)}\!,\!\psi^{(\hat{t}\,)} )$ is a convex optimization problem, its optimal solution can be obtained using standard convex optimization solvers (e.g., SDPT3). The complete procedure for transmit power optimization is summarized in Algorithm~\ref{alg:3}.
\begin{algorithm}[t]
\caption{Dinkelbach and quadratic method for $\mathcal{P}$ in $\mathscr{P}_4$}
\label{alg:3}
\begin{algorithmic}
\State \textbf{Input:} $\{ \mathbf{w}_\ell, \mathcal{K}_\ell \}_{\ell
\in\mathcal{L}}$, $T_{\mathrm{D}}$, and $ T_{\mathrm{Q}}$ 
\State \textbf{Initialize:} $\check{t}=0$ and $p_{\ell,k}^{(\check{t})} = P_\mathrm{max}/K,\,k\in\mathcal{K}_\ell,\ell\in\mathcal{L}$
\Repeat
    \State Update $\chi^{(\check{t}\,)}$ using \eqref{eq:49}
    \State Set $\hat{t}=0$
    \Repeat
        \State Update $\psi_{k}^{(\hat{t}\,)}$ using \eqref{eq:51}
        \State Solve $\ddot{\mathscr{P}}_4 ( {\check{t},\hat{t}}\,)$ for $\mathcal{P}$
        \State $\hat{t}\gets \hat{t}+1$
    \Until convergence of $\psi_k^{(\hat{t}\,)}$ or $\hat{t} < T_{\mathrm{Q}}$
    \State $\check{t}\gets \check{t}+1$
\Until convergence of $\chi^{(\check{t}\,)}$ or $\check{t} < T_{\mathrm{D}}$
\State \textbf{return:} $\mathcal{P}^\star \leftarrow \mathcal{P}^{(\hat{t}\,)}$
\end{algorithmic}
\end{algorithm}
Finally, the original problem $\mathscr{P}_0$ is solved by combining the distributed precoder design with the centralized two-stage algorithm. The overall procedure is summarized in Algorithm~\ref{alg:4}.
\begin{algorithm}[t]
\caption{Overall Algorithm for $\mathcal{V}_0$ in $\mathscr{P}_0$ }
\label{alg:4}
\begin{algorithmic}
\State \textbf{Input:} sCSI and traffic demand $\{R^{\mathrm{(dm)}}_{k} \}_{k\in\mathcal{K}}$ 
\State Update $\{ \mathbf{r}_\ell,\mathcal{K}_\ell \}_{\ell\in\mathcal{L}} $ using Algorithm 2
\State Update $ \mathcal{P}$ using Algorithm 3
\State Update $\{ \mathbf{F}_{\mathrm{RF},\ell}, \mathbf{F}_{\mathrm{BB},\ell,k} \}_{k\in\mathcal{K_\ell},\ell\in\mathcal{L}}$  using \eqref{eq:22} and \eqref{eq:24}
\State \textbf{return:} $\mathcal{V}^\star_0$
\end{algorithmic}
\end{algorithm}

\begin{figure*}[!t]
\begin{align}
R_k^{\mathrm{ip}}=\log_2\!\left(\!
1 \!+\!
\frac{
|\check{\eta}|^2
\Big|
\sum\limits_{\ell\in\mathcal L_k}
\sqrt{\upsilon_{\ell,k}\kappa_{\ell,k}p_{\ell,k}}
\mathbf a_{\ell,k}^{\mathrm H}\mathbf x_{\ell,k}^{\varsigma}
\Big|^2
}{
\sum\limits_{u\in\mathcal K}\!
\sum\limits_{\ell\in\mathcal L_u}\!
\upsilon_{\ell,k}p_{\ell,u}
\Big|\mathbf a_{\ell,k}^{\mathrm H}\mathbf x_{\ell,u}^{\varsigma}\Big|^2\!\!
\left(1\!+\!\kappa_{\ell,k}\left(1\!-\!|\check{\eta}|^2\right)\right)
\!+\!
|\check{\eta}|^2\!\!\!\!\!
\sum\limits_{u\in\mathcal K\setminus{\{k\}}}\!
\Big|\!
\sum\limits_{\ell\in\mathcal L_u}\!\!
\sqrt{\upsilon_{\ell,k}\kappa_{\ell,k}p_{\ell,u}}
\mathbf a_{\ell,k}^{\mathrm H}\mathbf x_{\ell,u}^{\varsigma}
\Big|^2
\!\!\!+\!
Q_k\!+\!\sigma^2
}
\right)
\label{eq:ip_rate} \tag{58}
\end{align}
\hrule
\vspace{-0.4cm}
\end{figure*} 

{\vspace{-0.4cm}\color{black}\subsection{Robust Design under Imperfect Pre-Compensation} \label{section5C}

While the preceding channel model assumes perfect pre-compensation of the phase term due to the satellite Doppler shift and minimum propagation delay, practical pre-compensation could be imperfect because of sCSI uncertainty and hardware impairments. To capture this imperfection, the residual compensation error is modeled as a multiplicative phase uncertainty applied to the perfect channel. Specifically, when the stream intended for the $u$-th UE is observed at the $k$-th UE, the channel from the $\ell$-th satellite is expressed as
\begin{align}
\mathbf h_{\ell,k,u}^{\mathrm{ip}}
&=
\mathrm{e}^{-j\Delta_{\ell,k,u}}\mathbf h_{\ell,k} \nonumber\\
&=
\mathrm{e}^{-j\Delta_{\ell,k,u}}
\sqrt{\upsilon_{\ell,k}}
\left(\sqrt{\kappa_{\ell,k}}+\alpha_{\ell,k}\right)
\mathbf a_{\ell,k},
\label{eq:ip_channel} 
\end{align}
where $\!\Delta_{\ell,k,u}\!$ denotes the residual compensation error. This model provides a tractable representation of the residual phase that remains after nominal pre-compensation and is applied to both the intended link and inter-UE interference links.

For analytical tractability, the residual phase uncertainty is characterized by its first-order moment as
\begin{align}
\check{\eta} \triangleq \mathbb E\left[\mathrm{e}^{-j\Delta_{\ell,k,u}}\right],
\qquad
|\check{\eta}|\leq 1,
\label{eq:phase_moment} 
\end{align}
Here, $|\check{\eta}| \!=\! 1$ corresponds to the case of the perfect channel, whereas smaller values of $|\check{\eta}|$ indicate larger residual phase uncertainty. Under this model, the aggregate effective channel coefficient $\dot{g}_{k,u}$ in \eqref{eq:new11} is extended to account for imperfect pre-compensation and is denoted by
\begin{align}
\dot{g}_{k,u}^{\mathrm{ip}} = 
\sum_{\ell\in\mathcal L_u}
(\mathbf h_{\ell,k,u}^{\mathrm{ip}})^{\mathrm H}
\mathbf x_{\ell,u}^{\varsigma}
\sqrt{p_{\ell,u}} .
\label{eq:ip_g}
\end{align}
Assuming that the residual compensation errors and the small-scale fading across satellite–UE links are mutually independent, the mean and variance of $\dot{g}_{k,u}^{\mathrm{ip}}$ are defined as
\begin{align}
&\mathbb E \big\{ \dot{g}_{k,u}^{\mathrm{ip}}\big\}
\! = \!
\check{\eta}
\sum_{\ell\in\mathcal L_u}
\sqrt{\upsilon_{\ell,k}\kappa_{\ell,k}p_{\ell,u}}
\mathbf a_{\ell,k}^{\mathrm H}\mathbf x_{\ell,u}^{\varsigma},
\label{eq:ip_mean}\\
&\mathbb{V}\big\{\dot{g}_{k,u}^{\mathrm{ip}}\big\}
\!=\!\!\!
\sum_{\ell\in\mathcal L_u} \!
\upsilon_{\ell,k}p_{\ell,u} \!
\left|\mathbf a_{\ell,k}^{\mathrm H}\mathbf x_{\ell,u}^{\varsigma}\right|^2\!\!\!
\left(1+\kappa_{\ell,k}\left(1-|\check{\eta}|^2\right)\right).
\label{eq:ip_var}
\end{align}

By substituting \eqref{eq:ip_mean} and \eqref{eq:ip_var} into the original UatF-based rate expression in \eqref{eq:12}, the achievable rate for the case of the imperfectly pre-compensation channel is derived, as presented in \eqref{eq:ip_rate} at the top of this page.
Here,
$Q_k
\!\triangleq\!
 \sum_{\ell\in\mathcal L} \gamma_{\ell,k}
 \operatorname{tr}(
(\mathbf F_{\mathrm{RF},\ell}^{\varsigma})^{\mathrm H}
\mathbf a_{\ell,k}\mathbf a_{\ell,k}^{\mathrm H}
\mathbf F_{\mathrm{RF},\ell}^{\varsigma}
\mathbf R_{\epsilon_\ell}
)$
denotes the received DAC quantization-noise power and is identical to that in the original UatF-based rate expression, regardless of pre-compensation errors.

Since imperfect pre-compensation only modifies the rate expression, the proposed two-stage framework remains applicable by replacing $R_k$ with $R_k^{\mathrm{ip}}$ in the original optimization problem. Moreover, when $\check{\eta} = 1$, Eq. \eqref{eq:ip_rate} reduces to the original rate formula, confirming that the robust design is a direct generalization of the original one.
It is also worth noting that the robust design does not require knowledge of the full distribution of the residual compensation error. Knowing only the first-order phase moment $\check{\eta}$ suffices to capture the impact of residual errors in the UatF-based rate expression.}

\subsection{Complexity Analysis} \label{section4D}
\subsubsection{Stage 1 (Modified CE)}
Each CE sample entails precoder computation and a bisection search.
The former is dominated by RF precoder generation and ZF baseband precoding. 
For the $\ell$-th satellite, RF precoding scales as $\mathcal{O}(N_\mathrm{t} |\mathcal{R}_\ell|)$, while ZF baseband precoding scales as $\mathcal{O}(|\mathcal{R}_\ell|^3)$ due to matrix inversion over the active RF chain dimension. The bisection search requires $T_\mathrm{B}= \left\lceil\log_2\left(P_\mathrm{max}/\varepsilon_\mathrm{B}\right)\right\rceil$ iterations to meet tolerance $\varepsilon_\mathrm{B}$, and each iteration evaluates the rate-related terms over $K$ UEs, yielding $\mathcal{O}(T_\mathrm{B}K)$. Furthermore, at each CE iteration, the sample sorting and PMF update steps incur computational costs of $\mathcal{O}(N_\mathrm{s}\log N_\mathrm{s})$ and $\mathcal{O}(N_\mathrm{elite}L)$, respectively. Since the modified CE procedure evaluates $N_\mathrm{s}$ samples per iteration over at most $T_{\mathrm{CE}}$ iterations, the complexity of Stage~1 is
\( 
\mathcal{O} (T_{\mathrm{CE}}[N_\mathrm{s} (\sum_{\ell\in\mathcal{L}}(N_\mathrm{t}|\mathcal{R}_\ell|+|\mathcal{R}_\ell|^3)+T_\mathrm{B}K ) + N_\mathrm{s} \log N_\mathrm{s} + N_\mathrm{elite} L  ] ). \)
Using $|\mathcal{R}_\ell| \le N_{\mathrm{RF}}$ and noting that the PMF update cost can be included in the sample evaluation cost, the complexity of Stage~1 is reformulated as 
\setcounter{equation}{58}
\begin{align}
    \mathcal{O}(T_{\mathrm{CE}}[ N_\mathrm{s}(L(N_\mathrm{t}N_{\mathrm{RF}}+N_{\mathrm{RF}}^3)+T_\mathrm{B}K)+ N_\mathrm{s} \log N_\mathrm{s}]). \label{eq:complexity1} 
\end{align} 

\vspace{-1em}\subsubsection{Stage 2 (Dinkelbach method and quadratic transform)}
The convex subproblem \eqref{eq:51eq} has $J_\mathrm{P}$ optimization variables, where  
$J_\mathrm{P} = \sum_{\ell\in\mathcal{L}}|\mathcal{K}_\ell| + K + 1$.  
When employing the interior-point method with precision $\epsilon_{\mathrm{ip}}$, each iteration incurs a computational complexity of $\mathcal{O}\!\left(J_\mathrm{P}^{3.5}\log(1/\epsilon_{\mathrm{ip}})\right)$.  
Considering at most $T_\mathrm{D}$ Dinkelbach and $T_\mathrm{Q}$ quadratic transform iterations, the total Stage~2 complexity is  
\begin{align}
    \mathcal{O}\!\left(T_\mathrm{D} T_\mathrm{Q}\, J_\mathrm{P}^{3.5}\log(1/\epsilon_{\mathrm{ip}})\right). 
\end{align}

\subsubsection{Overall Complexity and Scalability}
The overall computational complexity is given by the sum of the computational complexities of Stage~1 and Stage~2: $\mathcal{O}(T_{\mathrm{CE}}[ N_\mathrm{s}(L(N_\mathrm{t}N_{\mathrm{RF}} \!+\! N_{\mathrm{RF}}^3) \!+ \!T_\mathrm{B}K)\!+\! N_\mathrm{s} \log N_\mathrm{s}] \! + \!
T_\mathrm{D}T_\mathrm{Q}J_\mathrm{P}^{3.5}\log(1/\epsilon_{\mathrm{ip}})
).$ 
This overall expression describes how the complexity of the proposed algorithm scales with key system parameters, specifically the numbers of satellites, RF chains, and UEs, in the worst case. In Stage~1, the dominant term is $L(N_{\mathrm{t}}N_{\mathrm{RF}}+N_{\mathrm{RF}}^3)$. In Stage~2, the complexity is mainly governed by the effective power allocation dimension $J_{\mathrm{P}}=\sum_{\ell\in\mathcal{L}}|\mathcal{K}_\ell|+K+1$, which admits the worst-case upper bound $\mathcal{O}(LN_{\mathrm{RF}}+K)$. Hence, the primary contributor to the overall computational complexity is closely related to the product of the numbers of satellites and RF chains. Nevertheless, the proposed algorithm can limit the number of active transmission links, thereby potentially keeping the resulting computational burden practically manageable.

\section{Numerical Results} \label{section5}
\subsection{Simulation Environment}
The simulation parameters used in the numerical evaluation are summarized in Table~\ref{tab:simulation}. A uniform traffic demand distribution is assumed for simplicity while it is straightforward to accommodate spatio-temporal traffic variations. The channel and hardware parameters are selected based on relevant standards and prior studies, and unless otherwise specified, all values listed in the table are used throughout the simulations.

\begin{table}[t]
\centering
\caption{Summary of simulation parameters}
\label{tab:simulation}
\renewcommand{\arraystretch}{1.15}
\begin{tabular}{p{0.42\columnwidth}|p{0.43\columnwidth}}
\Xhline{0.8pt}
Parameter & Value \\
\hline
\rowcolor{SectionGray}
\multicolumn{2}{l}{System and channel parameters} \\
\hline
\rowcolor{LightGray}
$L$, $K$, Service area
& $5$, $10$, $700\,\mathrm{km}\!\times\!700\,\mathrm{km}$\cite{kimcellfree} \\
Satellite altitude; off-nadir angle
& $500\,\mathrm{km}$\cite{3gpp38811}; $35^\circ$\cite{kimcellfree}  \\
\rowcolor{LightGray}
UPA size, $N_{\mathrm{RF}}$
& $16\times16$\cite{LiYou_1} , $4$\cite{9238423} \\
$f_\mathrm{c}$; $G^{\mathrm{sat}}$; $G^{\mathrm{ue}}$
& $2\,\mathrm{GHz}$\cite{LiYou_1}; $8\,\mathrm{dBi}$\cite{ratematching}; $0\,\mathrm{dBi}$\cite{3gpp38811} \\
\rowcolor{LightGray}
$\kappa_{\ell,k}$; $R_k^{\mathrm{(dm)}}$
& $12\,\mathrm{dB}$~\cite{3gpp38811}; $\mathcal{U}_{[1.5,3]}$ $\mathrm{bps/Hz}$\cite{ratematching}\\
\hline
\rowcolor{SectionGray}
\multicolumn{2}{l}{Hardware and power consumption parameters} \\
\hline
$b_{\mathrm{DAC}}$, $b_{\mathrm{PS}}$; $P_{\mathrm{max}}$, $\sigma^2$
& $6$, $3$ bits\cite{Choi2022}; $40$, $10^{-11}\,\mathrm{dBm}$\cite{kimcellfree} \\
\rowcolor{LightGray}
$P_{\mathrm{PS}}\,(b_{\mathrm{PS}}=1,2,3,4)$
& $(5,\,10,\,15,\,21.6)\,\mathrm{mW}$ \cite{11078749,8333733} \\
$P_{\mathrm{LP}}$, $P_{\mathrm{M}}$, $P_{\mathrm{LO}}$, $P_{\mathrm{H}}$
& $14$, $0.3$. $22.5$, $3\,\mathrm{mW}$\cite{8333733} \\
\rowcolor{LightGray}
$\eta_{\mathrm{PA}}$, $f_\mathrm{s}$
& $0.27$, $10^8$ samples/s~\cite{Choi2022} \\
\hline
\rowcolor{SectionGray}
\multicolumn{2}{l}{Algorithm parameters} \\
\hline
$N_\mathrm{s}$, $\lambda$; $\digamma$, $\varepsilon_{\mathrm{B}}$, $\vartheta_\mathrm{th}$
& $10^6$, $0.1$, $0.5$, $10^{-6}$, $7^\circ$ \\
\rowcolor{LightGray}
$T_{\mathrm{B}}$, $T_{\mathrm{CE}}$, $T_{\mathrm{D}}$, $T_{\mathrm{Q}}$
& $100$, $20$, $5$, $10$ \\
\Xhline{0.8pt}
\end{tabular}
\end{table}

\subsection{Benchmarks}
For comparison, four benchmark schemes are considered:
\begin{enumerate}[label=\textit{(\roman*)}]
    \item \textbf{CE-NC}: a non-cooperative CE scheme in which each UE is associated with at most one satellite via the CE procedure \cite{CEbj};
    \item \textbf{DD-G}: a greedy scheme in which each UE is first associated with its closest satellite \cite{9079921}, after which additional associations are established in descending order of traffic demand $R_k^{(\mathrm{dm})}$;
    \item \textbf{DA-PA}: a demand-aware power allocation scheme in which each satellite allocates $P_{\max}$ among its associated UEs in proportion to $R_k^{(\mathrm{dm})}$; and
    \item \textbf{GA-SDR}: a benchmark using a genetic algorithm (GA) in Stage~1~\cite{GA}, and semidefinite relaxation (SDR) in Stage~2~\cite{SDR}
\end{enumerate}
For CE-NC and DD-G, one RF chain per served UE is activated under the FC architecture, whereas all RF chains are activated under the PC architecture, which exploits the array gain of all subarrays while still incurring lower circuit power than the FC architecture. GA--SDR is included as a representative optimization-based benchmark to assess the performance of the proposed framework under comparable evaluation conditions. For fairness, all benchmark components not explicitly modified above are implemented identically to those of the proposed algorithm.

\subsection{Performance Comparison with Benchmark Schemes}

\begin{figure}[t]
    \centering\includegraphics[width=0.9\columnwidth]{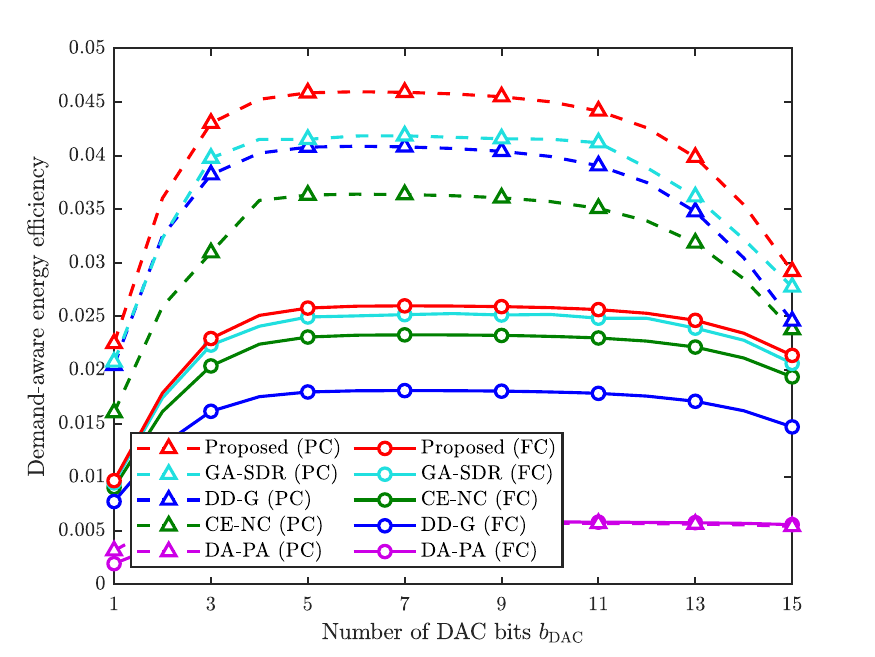}
    \caption{Demand-aware EE as a function of DAC resolution
    $b_{\mathrm{DAC}}$.}\label{fig:3}
\end{figure}
Fig.~\ref{fig:3} illustrates how the demand-aware EE of the proposed framework and the four benchmark schemes varies with the DAC resolution $b_{\mathrm{DAC}}$. For both PC and FC architectures, the proposed framework attains the highest demand-aware EE over the entire range of $b_{\mathrm{DAC}}$, with its superiority being especially pronounced at moderate DAC resolutions. The PC architecture consistently outperforms FC in demand-aware EE because it uses fewer PS elements per RF chain.
As $b_{\mathrm{DAC}}$ increases from one to about five bits, demand-aware EE improves because the reduced quantization distortion outweighs the extra DAC power. Beyond this, further increases in $b_{\mathrm{DAC}}$ yield diminishing gains as the circuit power dominates. This behavior shows that the proposed framework allows the use of moderate-resolution DACs while nearly achieving the optimal demand-aware EE performance.
The gap between the proposed framework and DA-PA demonstrates the significance of transmit power optimization and indicates that simply using the maximum transmit power is not an appropriate strategy for EE-based transmission.

\begin{figure}[!t]
    \centering\includegraphics[width=0.9\columnwidth]{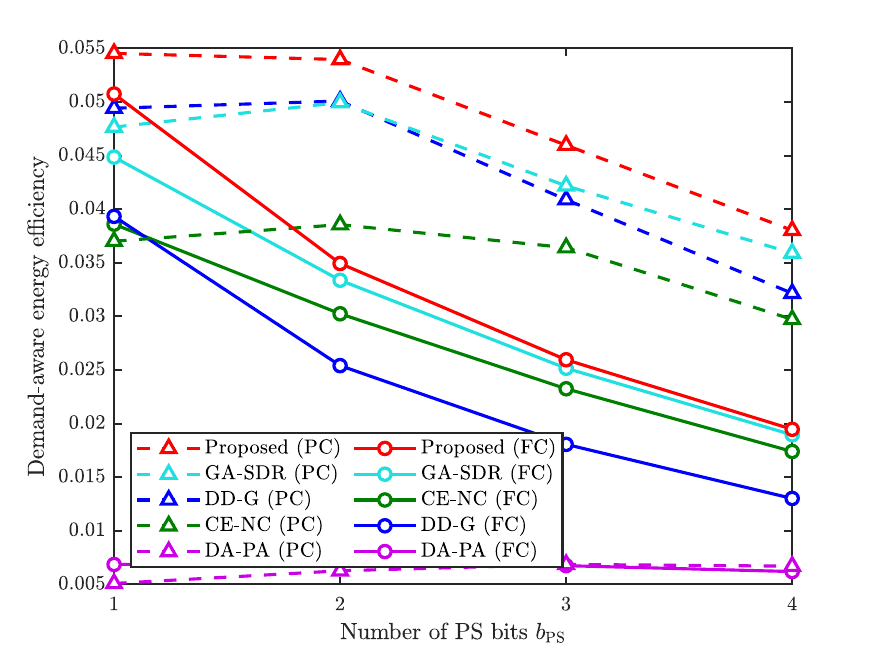}
    \caption{Demand-aware EE as a function of PS resolution
    $b_{\mathrm{PS}}$.}\label{fig:4}
\end{figure}
Fig.~\ref{fig:4} depicts the impact of the PS resolution $b_{\mathrm{PS}}$ on the demand-aware EE. For all considered values of $b_{\mathrm{PS}}$, the proposed framework outperforms the benchmark methods. In both architectures, the tradeoff between array gain and PS power consumption yields the optimal operating point at low resolution: for the proposed framework, the highest demand-aware EE is achieved at $b_{\mathrm{PS}}\!=\!1$, while increasing $b_{\mathrm{PS}}$ degrades it because the additional PS power outweighs the incremental improvement in array gain. Overall, the results indicate that coarse PS quantization is sufficient to capture the main benefits of HPC in LEO satellite networks.

\begin{figure}[t]
    \centering\includegraphics[width=0.9\columnwidth]{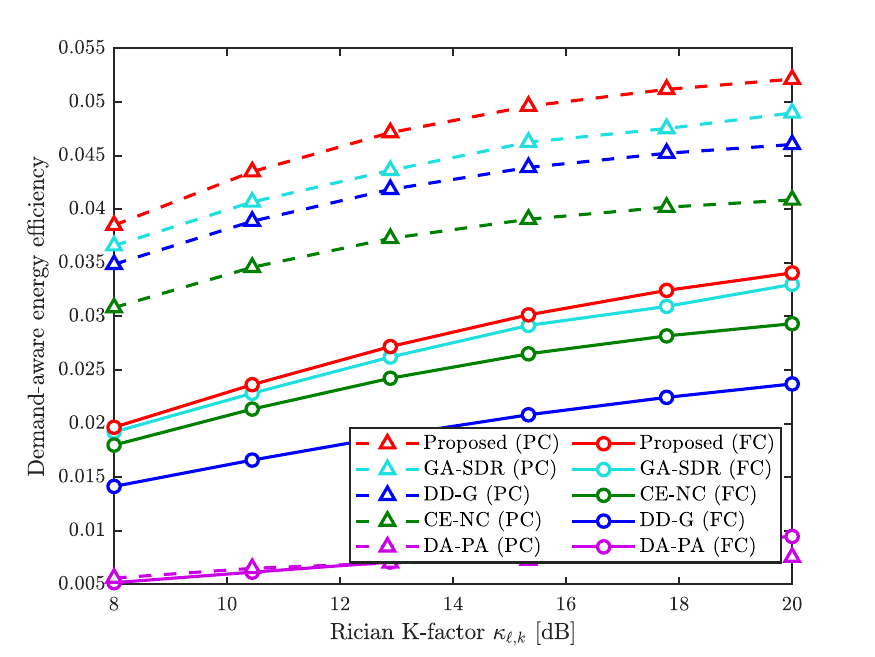}
    \caption{Demand-aware EE as a function of Rician K-factor $\kappa_{\ell,k}$.}\label{fig:5}
\end{figure}
Fig.~\ref{fig:5} depicts how the demand-aware EE varies with the Rician K-factor. As $\kappa_{\ell,k}$ increases, the channel becomes more LoS-dominant, which enhances channel hardening and reduces precoding uncertainty\cite{kimcellfree}. Consequently, all schemes benefit from reduced small-scale fading and exhibit a gradual increase in demand-aware EE. Over the entire range of $\kappa_{\ell,k}$, the proposed framework achieves the highest EE. The performance gain of the proposed framework is significant in the high-K regime, where the deterministic LoS component dominates and the sCSI-based beam alignment is most effective.

\begin{figure}[!t]
    \centering\includegraphics[width=0.9\columnwidth]{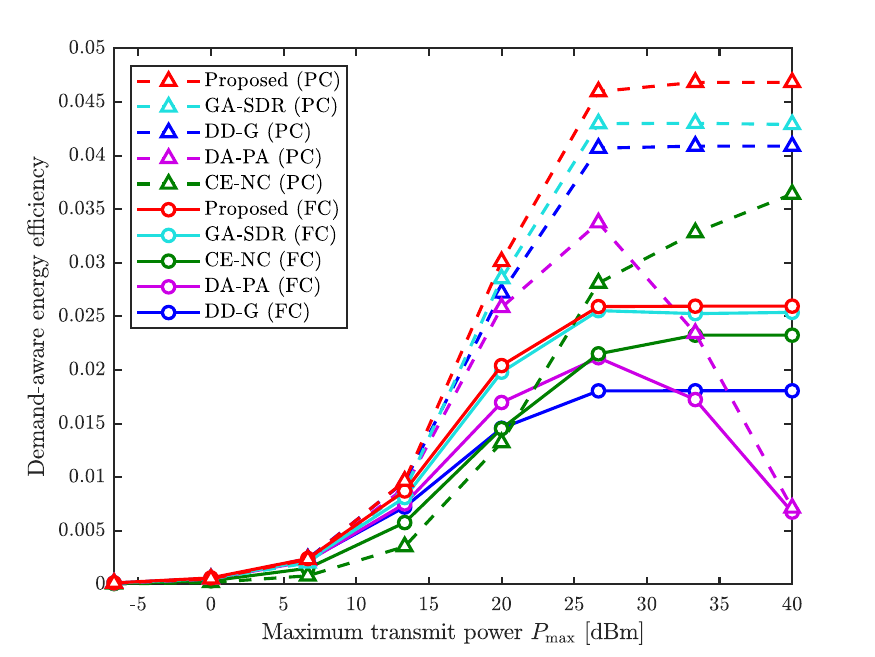}
    \caption{Demand-aware EE as a function of maximum transmit power $P_{\mathrm{max}}$.}
    \label{fig:6}
\end{figure}
Fig.~\ref{fig:6} illustrates the demand-aware EE as a function of the maximum transmit power $P_{\max}$. In the low-power region, all schemes are power-limited and therefore yield similarly low EE. As $P_{\max}$ moves into the moderate-power region, the proposed framework attains a sharper rise in EE by optimizing RF chain activation, UE association, and power allocation. Over the entire range of $P_{\max}$, the proposed framework consistently outperforms all benchmark schemes. This performance gap becomes more pronounced as the maximum transmit power increases, due to avoiding unnecessary transmit power and RF chain activation, while adapting the operating point according to the tradeoff between rate improvement and additional overall power consumption.

\begin{figure}[!t]
    \centering\includegraphics[width=0.9\columnwidth]{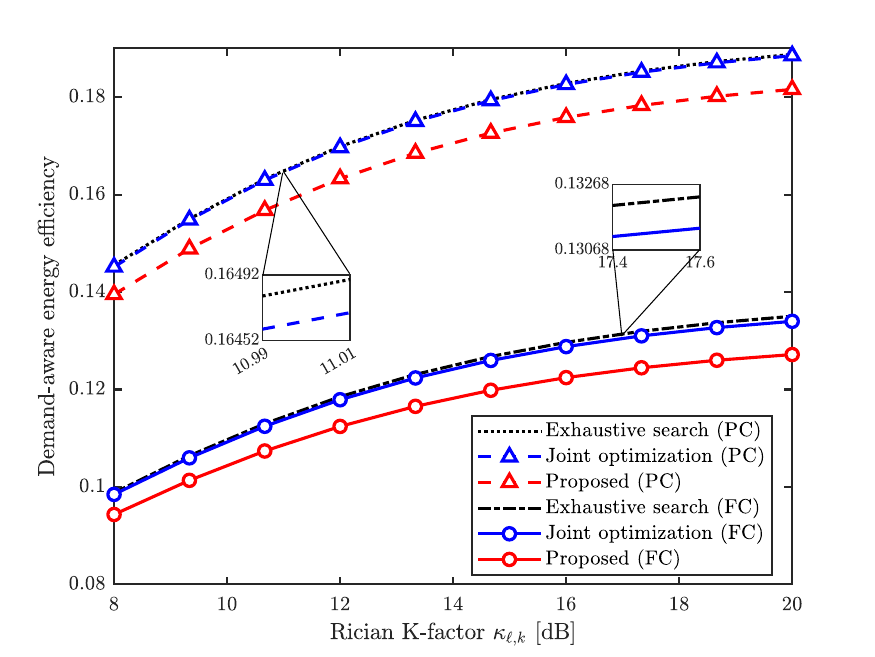}
    \caption{Average demand-aware EE as a function of the Rician K-factor over 10 different deployments.}
    \label{fig:R1_1_mean}
\end{figure}
To evaluate the performance loss caused by the proposed search-space reduction and two-stage decomposition, we compare the proposed algorithm with two additional upper-bound benchmarks: joint optimization and exhaustive search. The joint optimization benchmark simultaneously selects RF chain activation, UE association, and transmit power allocation by solving the transmit power allocation problem for each CE sample. The exhaustive search benchmark evaluates all RF chain activation and UE association combinations over the entire off-nadir feasible region, i.e., without restricting UE selection to the candidate UE set $\breve{\mathcal{K}}_\ell$, and then performs transmit power optimization. Since both upper-bound benchmarks are computationally prohibitive at the original network size, we consider a reduced setup with $L=2$, $N_{\mathrm{RF}}=3$, and $K=4$, and average the results over 10 independent deployments.
As shown in Fig.~\ref{fig:R1_1_mean}, the proposed algorithm maintains a limited and consistent performance gap with respect to both upper-bound benchmarks. In addition, the gap between joint optimization and exhaustive search is marginal, indicating that the candidate UE set $\breve{\mathcal{K}}_\ell$ preserves the dominant UE association candidates while substantially reducing the search space. Therefore, the proposed algorithm achieves a favorable performance–complexity trade-off by retaining most of the performance gain of joint optimization while avoiding its prohibitive computational burden.

\begin{figure}[!t]
    \centering\includegraphics[width=0.9\columnwidth]{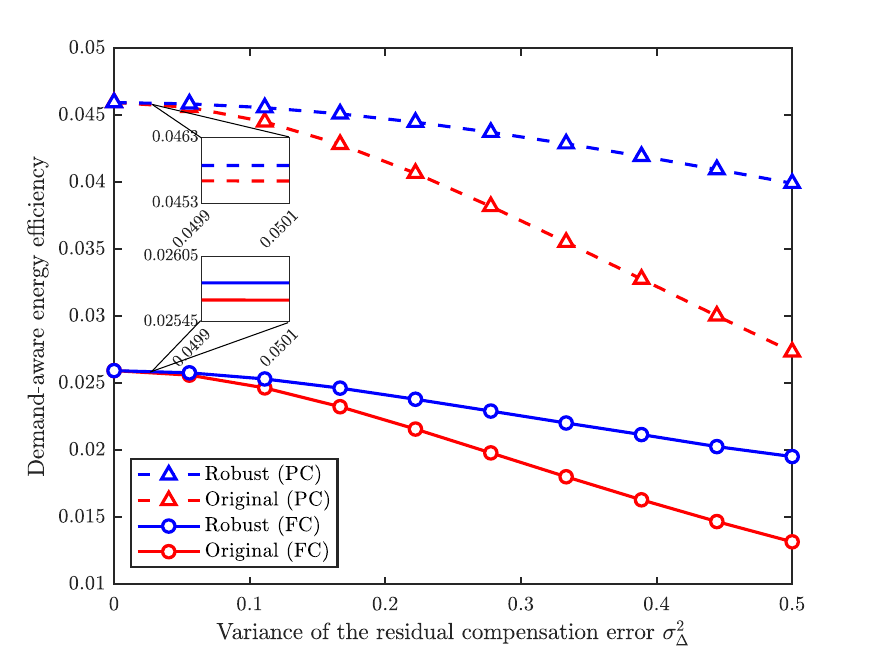}
    \caption{Demand-aware EE as a function of the variance of the residual compensation error.}
    \label{fig:compensation_error}
\end{figure}
\subsection{Performance Analysis under Imperfect Pre-Compensation}
To analyze the impact of imperfect pre-compensation, the residual compensation error $\Delta_{\ell,k,u}$ is modeled as a zero-mean, truncated Gaussian random variable on $[-\pi, \pi]$ with variance $\sigma_{\Delta}^{2}$.
Fig.~\ref{fig:compensation_error} compares the original and robust designs under imperfect pre-compensation as a function of $\sigma_{\Delta}^{2}$. Both exhibit gradual degradation for small $\sigma_{\Delta}^{2}$ because the UatF-based rate depends on the mean and variance of the aggregate effective channel rather than instantaneous phase errors.
As $\sigma_{\Delta}^{2}$ increases, the performance gap widens, showing that accounting for residual compensation errors in the optimization is crucial for robust cooperative transmission. The robust design provides a larger gain in the PC architecture than in the FC architecture because the PC architecture typically has more satellite–UE links and is therefore more sensitive to pre-compensation errors.

\subsection{Demand Satisfaction and Efficiency Tradeoff Analysis}
\begin{figure}[t]
  \centering
  \includegraphics[width=0.9\linewidth]{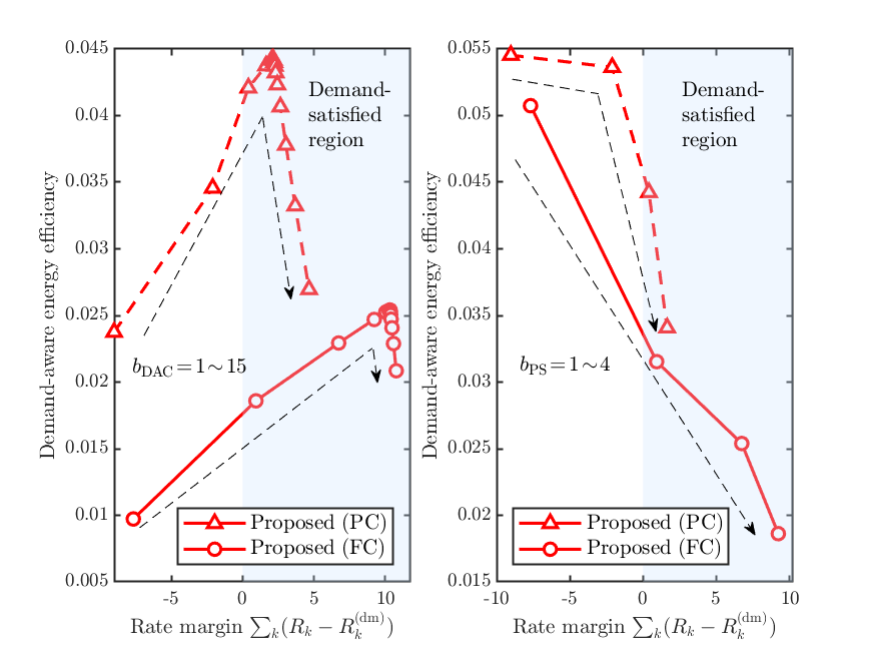}
  \caption{Demand-aware EE as a function of rate margin: impact of DAC resolution $b_{\mathrm{DAC}}$ (left) and PS resolution $b_{\mathrm{PS}}$ (right).}
  \label{fig:7}
\end{figure}
Fig.~\ref{fig:7} illustrates the tradeoff between demand-aware EE and the rate margin $\sum_{k} (R_{k} - R_{k}^{(\mathrm{dm})})$ for the proposed framework under both FC and PC architectures. The shaded area indicates the demand-satisfied region. In the left subfigure, the operating points are generated by increasing the DAC resolution from $b_{\mathrm{DAC}} = 1$ to $15$ bits along the arrows. The demand-aware EE optimum can lie within the demand-satisfied region: once the demand constraint is met, further increasing the positive margin does not necessarily improve EE and can even reduce it due to the extra DAC circuit power.
Thus, for both FC and PC architectures, there exists a desirable operating point that satisfies the required demand while achieving demand-aware EE optimality.
In contrast, the right subfigure, with $b_{\mathrm{PS}}$ varying from 1 to 4 bits, shows a clear separation between demand satisfaction and the demand-aware EE optimum: the maximum demand-aware EE is reached at negative margins, and shifting operation into the shaded region pushes the system to lower-EE points, making it hard to simultaneously meet demand and attain the demand-aware EE optimum.
Viewed from a Pareto-front perspective with the objectives of \textit{larger margin} and \textit{higher demand-aware EE}, sweeping the PS resolution produces a Pareto front; thus, all $b_{\mathrm{PS}}$ operating points remain non-dominated. In contrast, the DAC-resolution sweep includes low-resolution points that are strictly dominated by higher-resolution settings, which improve both margin and energy efficiency. Thus, very coarse $b_{\mathrm{DAC}}$ values are undesirable when jointly balancing demand satisfaction and energy efficiency.

\subsection{Impact of System Parameters on Demand-Aware EE}
\begin{figure}[t]
  \centering
  \includegraphics[width=0.9\linewidth]{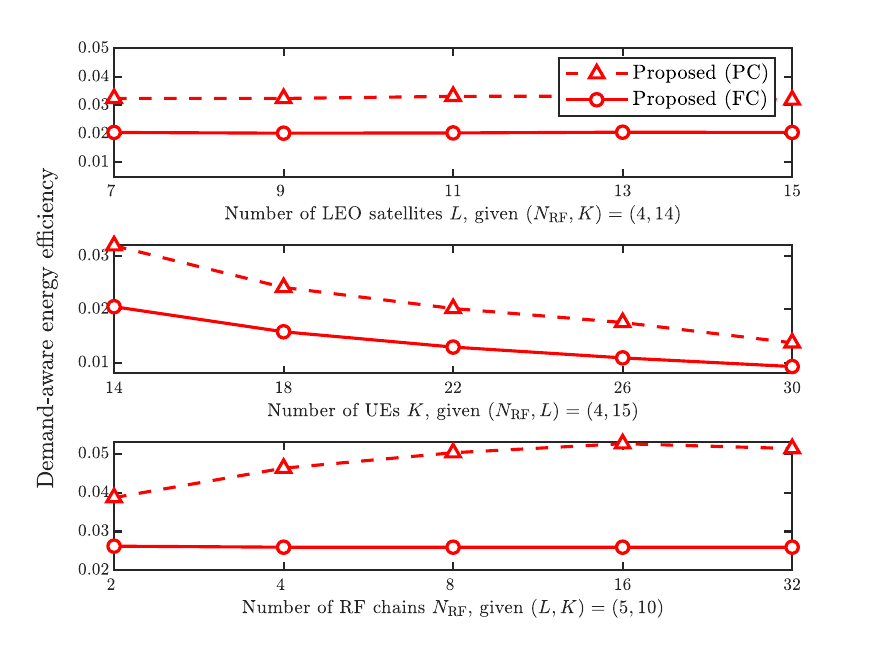}
  \caption{Demand-aware EE as a function of the number of LEO satellites $L$ (top), the number of UEs $K$ (middle), and the number of RF chains $N_{\mathrm{RF}}$ (bottom).}
  \label{fig:8}
\end{figure}
Fig.~\ref{fig:8} illustrates how the demand-aware EE behaves as key system parameters change. It remains nearly constant as $L$ increases because activating only the necessary RF chains prevents the addition of unnecessary links. In contrast, it decreases as $K$ grows because serving more UEs inevitably requires additional transmit power and circuit power, thereby increasing the total power consumption. As each additional RF chain incurs an architecture-dependent circuit power cost, the FC and PC architectures exhibit different trends as $N_{\mathrm{RF}}$ increases. In the PC architecture, the lower incremental cost allows more transmission links to be activated, creating a trade-off with the reduced array gain and leading to an optimal $N_{\mathrm{RF}}$. In contrast, the higher cost of the FC architecture limits redundant activation, keeping performance nearly constant.

\subsection{Validation of the Demand-Aware EE Objective}
\begin{figure}[t]
  \centering
  \includegraphics[width=0.9\linewidth]{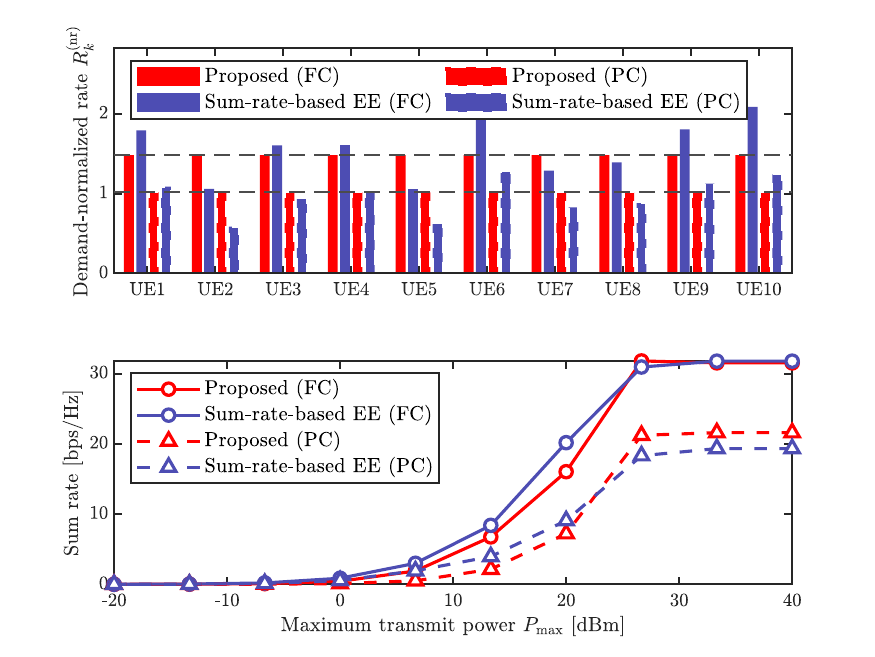}
  \caption{Comparison between the proposed demand-aware EE maximization and the sum-rate-based EE maximization under FC and PC architectures: per-UE demand-normalized rates $R^\mathrm{(nr)}_k$ (top) and sum rate as a function of maximum transmit power $P_{\max}$ (bottom).}
  \label{fig:9}
\end{figure}
Fig.~\ref{fig:9} compares the proposed demand-aware EE maximization with a conventional sum-rate-based EE maximization for both FC and PC architectures. For fairness, the sum-rate-based EE approach includes a connectivity constraint ensuring that each UE is served by at least one satellite.
In the upper subfigure, the proposed framework yields nearly identical demand-normalized rates across UEs, as indicated by their alignment with the black dashed reference line. In contrast, the sum-rate-based EE design exhibits much larger variability: certain UEs receive more resources than needed, while others are still underserved (i.e., $R^\mathrm{(nr)}_k \!\!<\!\! 1$), even when the proposed framework is capable of satisfying all UE requirements. 
Note that, since traffic demands are not imposed as hard constraints, the proposed framework remains feasible even when the aggregate traffic demands exceed the system capacity. Conversely, in resource-abundant regimes, any surplus rate can be leveraged for best-effort utilization of residual resources.
The lower subfigure shows the sum rate as a function of $P_{\max}$. In addition to explicitly accounting for UE-specific traffic demand, the proposed framework achieves a sum rate very close to that of the sum-rate-based EE benchmark over the entire $P_{\max}$ range for both FC and PC architectures. Hence, the proposed framework achieves nearly the same aggregate throughput as the sum-rate-oriented design while offering balanced, demand-proportional service across UEs.

\vspace{-0.2cm}{ \subsection{Convergence of the Proposed Two-Stage Algorithm}
\begin{figure}[t]
  \centering
  \includegraphics[width=0.9\linewidth]{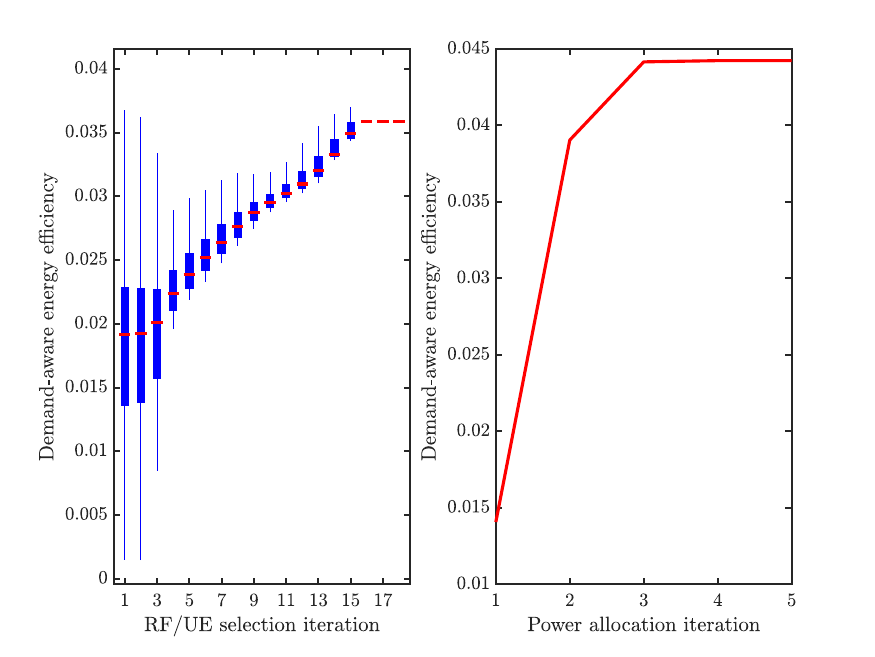}
  \caption{Convergence of the proposed two-stage algorithm: demand-aware EE as a function of CE iterations (left) and power-allocation iterations (right).}
  \label{fig:10}
\end{figure}
Fig.~\ref{fig:10} illustrates the convergence of the proposed two-stage algorithm. The left subfigure illustrates how demand-aware EE progresses over the iterations of the modified CE method, with the boxplots representing the objective values of the elites chosen at each step. The steadily increasing median, together with the eventual concentration in a single dominant configuration, confirms the stability and effectiveness of the CE-based selection of RF chain activation and UE association. The right subfigure shows the convergence of the transmit power allocation stage, where only a few Dinkelbach iterations are needed to reach the final EE. Overall, the results show that the proposed algorithm converges reliably.}\noindent

\vspace{0cm}\section{Conclusion} \label{section6}
This paper presents a demand-aware and energy-efficient cooperative transmission framework for multi-LEO satellite networks in environments with spatially non-uniform traffic demands. Satellites adopt an HPC architecture with RF chain activation and quantized DACs and PSs, and are organized into UE-centric clusters. Demand-aware EE is maximized by jointly optimizing HPC vectors, RF chain activation, UE association, and transmit power. Leveraging the single-AoD property of the satellite channel, distributed linear precoders are used in the proposed framework, and a centralized two-stage algorithm is employed: a modified CE method for RF chain activation and UE association, followed by power allocation using Dinkelbach’s method with a quadratic transform. \textcolor{black}{In addition, a robust design is developed to account for the effects of residual phase uncertainty under imperfect pre-compensation. Simulation results show that the proposed framework outperforms benchmark schemes and achieves demand-proportional rate allocation across UEs while maximizing EE.}

\appendices

\vspace{-0.1cm} 
\section{} \label{appendix:A} 

This appendix presents the derivation of $\beta_{\ell,k}$, the transmit power coefficient that incorporates both the information signal and the DAC quantization noise.
The DAC quantization noise covariance matrix is reformulated as
\(
    \mathbf R_{\boldsymbol{\epsilon}_{\ell}} = 
    q_\ell(1-q_\ell)
    \mathrm{diag}\!\left(
    \mathbb{E}\!\left\{
    \tilde{\mathbf v}_{\ell}\tilde{\mathbf v}_{\ell}^{\mathrm H}
    \right\}
    \right),
\)
where the simplification follows from the assumption of a uniform DAC resolution.
Starting from \eqref{eq:5} and using the definition of the reduced-dimensional baseband precoder, it follows that
\begin{align}
    \mathrm{diag}\!\left(
    \mathbb{E}\!\left\{
    \tilde{\mathbf v}_{\ell}\tilde{\mathbf v}_{\ell}^{\mathrm H}
    \right\}
    \right)
    =
    \sum_{k\in\mathcal K_\ell}
    p_{\ell,k}
    \mathrm{diag}\!\left(
    \mathbf f_{\mathrm{BB},\ell,k}
    \mathbf f_{\mathrm{BB},\ell,k}^{\mathrm H}
    \right).
    \label{eq:app_diag_vtilde}
\end{align}
The transmit power associated with the information signal of the $\ell$-th satellite is given by $\sum_{k\in\mathcal{K}_\ell} p_{\ell,k}$, because the effective HPC vector is normalized as in \eqref{eq:21d}. The transmit power contribution of the quantization noise after RF precoding can be expressed as
\begin{align}
    &\mathrm{tr}\!\left(
    \mathbb{E}\!\left\{
    \mathbf x_{\ell,\mathrm q}^{\varsigma}
    (\mathbf x_{\ell,\mathrm q}^{\varsigma})^{\mathrm H}
    \!\right\}\!
    \right)
    \!=\!
    \mathrm{tr}\!\left(\!
    (\mathbf F_{\mathrm{RF},\ell}^{\varsigma})^{\mathrm H}
    \mathbf F_{\mathrm{RF},\ell}^{\varsigma}
    \mathbf R_{\boldsymbol{\epsilon}_{\ell}}
    \!\right) \nonumber\\
    &\,\, \!=\!
    q_\ell(1-q_\ell) \!\!
    \sum_{k\in\mathcal K_\ell} \!
    p_{\ell,k}
    \mathrm{tr}\!\left( \!
    (\mathbf F_{\mathrm{RF},\ell}^{\varsigma})^{\mathrm H} 
    \mathbf F_{\mathrm{RF},\ell}^{\varsigma}
    \mathrm{diag}\!\left(
    \mathbf f_{\mathrm{BB},\ell,k}
    \mathbf f_{\mathrm{BB},\ell,k}^{\mathrm H}
    \right)\!
    \right).
    \label{eq:app_q_power_1}
\end{align}
Because each column of the RF precoder comprises $N_{\mathrm{PS}}^{\varsigma}$ phase shifters with constant-modulus coefficients, the diagonal entries of $(\mathbf F_{\mathrm{RF},\ell}^{\varsigma})^{\mathrm H}\mathbf F_{\mathrm{RF},\ell}^{\varsigma}$ are all equal to $N_{\mathrm{PS}}^{\varsigma}$. Therefore, $\mathrm{tr}( \mathbb{E}\!\{ \mathbf x_{\ell,\mathrm q}^{\varsigma} (\mathbf x_{\ell,\mathrm q}^{\varsigma})^{\mathrm H} \} ) \!=\! q_\ell(1-q_\ell) \sum_{k\in\mathcal K_\ell} p_{\ell,k} N_{\mathrm{PS}}^{\varsigma} \| \mathbf f_{\mathrm{BB},\ell,k} \|^2.$ Hence, the coefficient of $p_{\ell,k}$ that arises from DAC quantization noise and information signal can be expressed as
\begin{align}
    \beta_{\ell,k}
    \triangleq 1 +
    N_{\mathrm{PS}}^{\varsigma}
    q_\ell(1-q_\ell)
    \left\|
    \mathbf f_{\mathrm{BB},\ell,k}
    \right\|^2 .
    \label{eq:app_beta_def}
\end{align}

{\vspace{-0.4cm}\section{} \label{appendix:B}
Under the common power level optimization, the rate expression in \eqref{eq:13} can be rewritten in a simplified form as
\begin{align}
     R_k(P_{\mathrm{com}}) = \log_2 \left( 1 + \frac{a_k P_{\mathrm{com}}}{b_k P_{\mathrm{com}} + c_k} \right), \quad \forall k\in\mathcal{K}, \label{eq:52}
\end{align}
where \(a_k, b_k, \text{and } c_k (> 0)\) denote the desired signal gain, the effective
interference, and the noise, respectively.
The first derivative of~\eqref{eq:52} with respect to \(P_{\mathrm{com}}\) is positive, and the second derivative is negative for all \(P_{\mathrm{com}} \ge 0\). Hence, Eq.~\eqref{eq:52} is strictly increasing and concave in \(P_{\mathrm{com}}\).
Since the \(k\)-th UE demand-normalized rate in \eqref{eq:new20} is a positive scaling of \eqref{eq:52}, each \(R^\mathrm{(nr)}_k(P_{\mathrm{com}})\) is concave and strictly increasing in \(P_{\mathrm{com}}\), and their pointwise minimum in \eqref{eq:new21} is also concave and strictly increasing.

Based on \eqref{eq:31}, the sign of ${f}'$ is completely determined by \(N^\mathrm{(sg)}\), and differentiating \(N^\mathrm{(sg)}\) once more gives
\(
  (N^\mathrm{(sg)})'
  = (R^\mathrm{(nr)}_{\min})'' P_\mathrm{tot},
\)
since \(P_\mathrm{tot}''\) is zero. For notational simplicity, the common argument \(P_{\mathrm{com}}\) is omitted.
For all $P_\mathrm{com} \ge 0$ with $(R^\mathrm{(nr)}_{\min})'' < 0$ and  $P_\mathrm{tot} > 0$, it follows that $(N^\mathrm{(sg)})' < 0$ for all $P_\mathrm{com} \ge 0$. Therefore, $N^\mathrm{(sg)}$ is strictly decreasing, and, since a simple calculation shows that $N^\mathrm{(sg)}(0) > 0$, this implies that $N^\mathrm{(sg)}$ can have at most one zero in the feasible interval $[0, P_{\mathrm{eff}}]$.
Consequently, $f$ is unimodal over $[0, P_{\mathrm{eff}}]$, that is, it attains a single maximum in this interval.} 


%







\bibliographystyle{IEEEtran}
\bibliography{./IEEEabrv,./reference}%
\end{document}